\documentclass[useAMS,usenatbib]{mnras}
\usepackage{txfonts,graphicx,amssymb,subfigure,color}
\usepackage[normalem]{ulem}
\usepackage{natbib}

\usepackage[hang,flushmargin]{footmisc}

\newcommand\blue{\color{blue}}

\defcitealias{heesen_16a}{HD16}
\defcitealias{croston_14a}{CH14}

\begin{document}
\title[A low-frequency study of 3C~31 with LOFAR]{LOFAR reveals the giant: a
  low-frequency radio continuum study of the outflow in the nearby FR~I radio galaxy 3C~31}

\author[V.~Heesen et al.]{V.~Heesen,$^{1,2}$\thanks{E-mail: {\blue volker.heesen@hs.uni-hamburg.de}} J.~H.~Croston,$^{1,3}$ R.~Morganti,$^{4,5}$
  M.~J.~Hardcastle,$^{6}$ A.~J.~Stewart,$^{7}$ \newauthor P.~N.~Best,$^{8}$ J.~W.~Broderick,$^{4}$
  M.~Br{\"u}ggen,$^{2}$ G.~Brunetti,$^{9}$ K.~T.~Chy{\.z}y,$^{10}$
  J.~J.~Harwood,$^{4}$ \newauthor M.~Haverkorn,$^{11}$ K.~M.~Hess,$^{4,5}$ H.~T.~Intema,$^{12}$
M.~Jamrozy,$^{10}$ \newauthor M.~Kunert-Bajraszewska,$^{13}$ 
J.~P.~McKean,$^{4,14}$ E. Orr{\'u},$^{4}$ H.~J.~A.~R{\"o}ttgering,$^{12}$ 
\newauthor T.~W.~Shimwell,$^{12}$  A.~Shulevski,$^{4}$ G.~J.~White,$^{3,15}$  E.~M.~Wilcots,$^{16}$ and W.~L.~Williams$^{6}$\\
$^{1}$School of Physics and Astronomy, University of Southampton,
Southampton SO17 1BJ, UK\\
$^{2}$University of Hamburg, Hamburger Sternwarte, Gojenbergsweg 112, 21029 Hamburg, Germany\\
$^{3}$School of Physical Sciences, The Open University, Walton Hall,
Milton Keynes, MK6 7AA, UK\\
$^{4}$ASTRON, Netherlands Institute for Radio Astronomy, PO 2, 7990 AA,
Dwingeloo, The Netherlands\\
$^{5}$Kapteyn Astronomical Institute, University of Groningen, PO 800, 9700
AV, Groningen, The Netherlands\\
$^{6}$School of Physics, Astronomy and Mathematics, University of
Hertfordshire, Hatfield AL10 9AB, UK\\
$^{7}$University of Oxford, Department of Physics, Astrophysics, Denys
Wilkinson Building, Keble Road, Oxford OX1 3RH, UK\\
$^{8}$SUPA, Institute for Astronomy, Royal Observatory, Blackford Hill, Edinburgh, EH9 3HJ, UK\\
$^{9}$INAF/Istituto di Radioastronomia, via Gobetti 101, I-40129 Bologna,
Italy\\
$^{10}$Astronomical Observatory, Jagiellonian University, ul. Orla 171, 30-244 Krak\'ow, Poland\\
$^{11}$Department of Astrophysics / IMAPP, Radboud University Nijmegen, PO Box 9010, 6500 GL Nijmegen, The Netherlands\\
$^{12}$Leiden Observatory, Leiden University, PO Box 9513, 2300 RA Leiden, The Netherlands\\
$^{13}$Toru\'n Centre for Astronomy, Faculty of Physics, Astronomy and
Informatics, NCU, Grudziacka 5, 87-100 Toru\'n, Poland\\
$^{14}$Kapteyn Astronomical Institute, University of Groningen, P.O. Box 800, 9700 AV Groningen, The Netherlands\\
$^{15}$RAL Space, The Rutherford Appleton Laboratory, Chilton, Didcot,
Oxfordshire OX11 0NL, UK\\
$^{16}$Dept. of Astronomy, University of Wisconsin, Madison, 475 N.~Charter Street, Madison, WI, 53706, USA
}

\date{Accepted 2017 October 21. Received 2017 October 18; in original form 2016 May 20}

\maketitle

\begin{abstract}
  We present a deep, low-frequency radio continuum study of the nearby
  Fanaroff--Riley class I (FR~I) radio galaxy 3C~31 using a
  combination of LOw Frequency ARray (LOFAR; 30--85 and 115--178~MHz),
  Very Large Array (VLA; 290--420~MHz), Westerbork Synthesis Radio
  Telescope (WSRT; 609~MHz) and Giant Metre Radio Telescope (GMRT;
  615~MHz) observations. Our new LOFAR 145-MHz map shows that 3C~31
  has a largest physical size of $1.1$~Mpc in projection, which means 3C~31 now
  falls in the class of giant radio galaxies. We model the
  radio continuum intensities with advective cosmic-ray transport,
  evolving the cosmic-ray electron population and magnetic field
  strength in the tails as functions of distance to the nucleus. We
  find that if there is no in-situ particle acceleration in the tails,
  then decelerating flows are required that depend on radius $r$ as
  $\varv\propto r^{\beta}$ ($\beta\approx -1$). This then compensates for the strong adiabatic losses due to the
  lateral expansion of the tails. We are able to find self-consistent
  solutions in agreement with the
  entrainment model of Croston \& Hardcastle, where the magnetic field
  provides $\approx$$1/3$ of the pressure needed for equilibrium with the
  surrounding intra-cluster medium (ICM). We obtain an advective time-scale of
  $\approx$$190$~Myr, which, if equated to the source age, would require an average expansion Mach number ${\cal M} \approx 5$ over the source lifetime. Dynamical arguments suggest that instead, either the outer tail material does not represent the oldest jet plasma or else the particle ages are underestimated due to the effects of particle acceleration on large scales.
\end{abstract}


%%%%%%%%%%%%%%%%%%%%%%%%%%%%%%%%%%%%%%%%%%%%%%%%%%%%%%%%%%%%%%%%%%%%%%%%%%%%
\section{Introduction}
\label{sec:introduction}
The jets of low-luminosity radio galaxies \citep[class I of][hereafter FR~I]{fanaroff_74a} are thought to
be relativistic decelerating flows that emanate from active galactic nuclei
(AGN). Models fitted to observations \citep{laing_02a,laing_04a} and numerical simulations
\citep{perucho_07a,perucho_14a} have convincingly shown that jets decelerate on kpc-scales
and can be described as relativistic flows observing the conservation of
particles, energy and momentum. Advances in the modelling of FR~I outflows were made possible
by combining radio and X-ray observations, which constrain the
density, temperature and pressure without having to rely on
the assumption of energy equipartition between cosmic rays and the magnetic field \citep{burbidge_56a}. This has made it possible to show that the outflow deceleration
can be explained by entrainment of material, both from internal and external entrainment. The former refers to
the entrainment of material stemming from sources inside the jet volume, such
as stellar winds, and the latter to entrainment of material from the intra-cluster
medium (ICM) via
shearing instabilities in the jet--ICM boundary layer. The inclusion of entrainment also can explain why the jet pressures, as
derived from energy equipartition of the cosmic rays and the magnetic field,
are smaller than the surrounding pressure as measured from the hot, X-ray
emitting gas \citep{morganti_88a, croston_14a}. 

The modelling of the FR~I outflow evolution has so far fallen into
  two categories. Within the first few kpc ($\lessapprox$$10$~kpc) the outflow is
  relativistic and narrow, so that
  relativistic beaming effects affect the apparent jet surface brightness; modelling these processes in total intensity and polarization allows the speeds and inclination angles to be constrained \citep[e.g.][]{laing_02a}. This area will be referred
  to hereafter as the \emph{jet} region. Further away from the nucleus (10--100~kpc),
  the outflow widens significantly with widths of a few kpc to a few 10~kpc and becomes sub-relativistic. This part of the
  outflow is commonly known as the \emph{radio tails} (also called \emph{plumes}), and in these regions the assumption of
  pressure equilibrium with the surrounding 
  X-ray emitting ICM can be used to estimate properties of the tails such as speed and
  magnetic and particle pressures \citep[e.g.][hereafter \citetalias{croston_14a}]{croston_14a}.

  In this paper, we present low-frequency radio continuum observations
  of the nearby FR~I radio galaxy 3C~31. This source is particularly
  well suited for jet modelling studies, because it is large, bright and nearby, 
  and these properties of the source have enabled some of the most detailed studies of radio-galaxy physics
  to date, including the modelling of physical parameters such as
  magnetic field strength and outflow speed. We acquired LOw Frequency
  ARray \citep[LOFAR;][]{van_Haarlem_13a} observations between 30 and
  178~MHz, which we combine with new Very Large Array (VLA) and Giant
  Metrewave Radio Telescope (GMRT) observations and archival
  Westerbork Synthesis Radio Telescope (WSRT) data between 230 and
  615~MHz. This paper is organized as
  follows: Section~\ref{sec:current_understanding} reviews our
  knowledge of the outflow in 3C~31, before we describe our
  observations and data reduction methods in
  Section~\ref{sec:observations}. Section~\ref{sec:morphology_and_observed_spectrum} contains
  a description of the radio continuum morphology and the observed radio continuum spectrum of 3C~31. In
  Section~\ref{sec:cr_transport}, we investigate the transport of cosmic-ray
  electrons (CREs) in the radio tails, employing a quasi-1D model of pure
  advection. We discuss our results in Section~\ref{sec:discussion} and
  present a summary of our conclusions in
  Section~\ref{sec:conclusions}. Throughout the paper, we use a
  cosmology in which $H_0 =70~\rm km\,s^{-1}~Mpc^{-1}$,
  $\Omega_{\rm m}=0.3$ and $\Omega_\Lambda=0.7$. At the redshift of
  3C~31 \citep[$z = 0.0169$;][]{laing_02a}, this gives a luminosity
  distance of $D = 73.3~\rm Mpc$ and an angular scale of
  $0.344~\rm kpc\, arcsec^{-1}$. Spectral indices $\alpha$ are defined
  in the sense $S_{\nu}\propto \nu^{\alpha}$, where $S_{\nu}$ is the (spectral)
  flux density and $\nu$ is the observing frequency. Reported errors
  are $1\sigma$, except where otherwise noted. Throughout the paper,
  the equinox of the coordinates is J$2000.0$. Distances from the
  nucleus are measured along the tail flow direction,
  accounting for bends, and corrected for an inclination angle to the line of sight of
  $\vartheta=52\degr$ \citep{laing_02a}.

\subsection{Current understanding of the outflow in 3C~31}
\label{sec:current_understanding}
3C~31 is a moderately powerful FR~I radio galaxy with a 178-MHz
luminosity of $9\times 10^{23}~\rm W\,Hz^{-1}\,sr^{-1}$. Because it is
relatively nearby, it has been extensively studied in the radio.
The two jets show a significant asymmetry in the radio continuum intensity,
with the northern jet much brighter than the southern jet on kpc-scales
\citep{burch_77a,ekers_81a,van_breugel_82a, laing_02a,laing_08a}. This asymmetry is reflected in
the radio polarization as well \citep{burch_79a,fomalont_80a,laing_02a} and
can also be seen on pc-scales \citep{lara_97a}. The brighter northern jet has
an optical counterpart \citep{croston_03a}, which is also detected in X-ray 
\citep{hardcastle_02a} and in infrared emission \citep{lanz_11a}; the spectrum
is consistent with synchrotron emission ranging from radio to X-ray frequencies.
3C~31 belongs to a sub-class of FR~I galaxies with limb-darkened radio
tails where the spectra steepen with
increasing distance from the nucleus \citep{strom_83a,andernach_92a,parma_99a}. It is hosted
by the massive elliptical galaxy NGC~383, which displays a dusty disc \citep{martel_99a},
and is a member of the optical chain Arp~331 \citep{arp_66a}. NGC~383 is the
most massive galaxy in a group or poor cluster of galaxies, containing
approximately 20 members within a radius of 500~kpc \citep{ledlow_96a}. The
group is embedded in extended X-ray emission from the hot ICM \citep{morganti_88a,komossa_99a,hardcastle_02a}.

High-resolution radio observations have allowed detailed
modelling of the velocity field in the jets within 10~kpc from the
nucleus, resting on the assumption that the differences in the jet
brightness can be explained entirely by Doppler beaming and aberration
of approaching and receding flows. The modelling by \citet{laing_02a}
showed that the jets have an inclination angle to the line of sight of $\approx$$52\degr$
and the on-axis jet speed is $\varv/c\approx 0.9$ at 1~kpc from the
nucleus, decelerating to $\varv/c\approx 0.22$ at 12~kpc, with slower
speeds at the jet edges. \citet{laing_02b} used the kinematic models
of \citet{laing_02a} and combined them with the X-ray observations of
\citet{hardcastle_02a} to show that the structure of the jets and
their velocity field can be explained by the conservation laws as
derived by \citet{bicknell_94a}. \citetalias{croston_14a}
used an {\it XMM--Newton}-derived external pressure profile to extend
the modelling of entrainment out to a distance of 120~kpc from the nucleus. They showed that at this distance from the nucleus the external
pressure is approximately a factor of 10 higher than the pressure
derived on the assumption of equipartition of energy between
relativistic leptons and magnetic field. \citetalias{croston_14a} favoured a model in which
continuing entrainment of material on 50--100-kpc scales accounts for
this discrepancy.

The present work extends the previous studies of 3C~31
\citep{laing_02a,laing_02b,croston_14a} to the physical conditions at
distances exceeding 120~kpc from the nucleus. This not only serves as
a consistency check, by extending previous work to lower frequencies, but
can also address the question of whether the entrainment model of
\citetalias{croston_14a} can be extended to the outskirts of the tails. It is also an
opportunity to revisit the spectral ageing analysis of 3C~31,
previously carried out by \citet{burch_77a}, \citet{strom_83a} and
\citet{andernach_92a}. These works suggest flow velocities of a few
$1000~\rm km\,s^{-1}$, using the spectral break frequency and assuming
equipartition magnetic field strengths. In particular,
\citet{andernach_92a} found a constant advection speed of
$5000~\rm km\,s^{-1}$ in the southern tail and an accelerating flow in
the northern tail.

\begin{table}
\centering
\caption{Journal of the observations.\label{tab:observations}}
\begin{tabular}{lc}
\hline\hline
\multicolumn{2}{c}{--- LOFAR LBA ---}\\\hline
Observation IDs & L96535\\
Array configuration & LBA\verb|_|OUTER\\
Stations & 55 (42 core and 13 remote)\\
Integration time & 1~s\\
Observation date & 2013 Feb 3\\
Total on-source time & 10~h (1 scan of 10~h)\\
Correlations & $XX$, $XY$, $YX$, $YY$\\
Frequency setup & 30--87~MHz (mean 52~MHz)\\
Bandwidth & 48~MHz (244 sub-bands)\\
Bandwidth per sub-band & $195.3125$~kHz\\
Channels per sub-band & 64\\
Primary calibrator & 3C~48 (1 scan of 10~h)\\
\hline
\multicolumn{2}{c}{--- LOFAR HBA ---}\\\hline
Observation IDs & L86562--86647\\
Array configuration & HBA\verb|_|DUAL\verb|_|INNER\\
Stations & 61 (48 core and 13 remote)\\
Integration time & 1~s\\
Observation date & 2013 Feb 17\\
Total on-source time & 8~h (43 scans of 11~min)\\
Correlations & $XX$, $XY$, $YX$, $YY$\\
Frequency setup & 115--178~MHz (mean 145~MHz)\\
Bandwidth & 63~MHz (324 sub-bands)\\
Bandwidth per sub-band & $195.3125$~kHz\\
Channels per sub-band & 64\\
Primary calibrator & 3C~48 (43 scans of 2~min)\\
\hline
\multicolumn{2}{c}{--- VLA $P$ band ---}\\\hline
Observation ID & 13B-129\\
Array configuration & A-array / B-array / C-array\\
Integration time & 1~s\\
Observation date & 2014 Apr 7 / 2013 Dec 14 / 2014 Dec 2\\
Total on-source time & 12~h (4~h per array, scans of 30~min)\\
Correlations & $XX$, $XY$, $YX$, $YY$\\
Frequency setup & 224--480~MHz (mean 360~MHz)\\
Bandwidth & 256~MHz (16 sub-bands)\\
Bandwidth per sub-band & 16~MHz\\
Channels per sub-band & 128\\
Primary calibrator & 3C~48 (10~min)\\
Secondary calibrator & J0119+3210 (scans of 2~min)\\
\hline
\multicolumn{2}{c}{--- GMRT ---}\\\hline
Observation ID & 12KMH01\\
Array configuration & N/A\\
Integration time & 16~s\\
Observation date & 2007 Aug 17\\
Total on-source time & $5.5$~h (scans of 30~min)\\
Correlations & $RR$\\
Frequency setup & 594--626~MHz (mean 615~MHz)\\
Bandwidth & 32~MHz (1 sub-band)\\
Bandwidth per sub-band & 32~MHz\\
Channels per sub-band & 256\\
Primary calibrator & 3C~48 (10~min)\\
Secondary calibrator & J0119+3210 (scans of 5~min)\\
\hline
\end{tabular}
\end{table}

\section{Observations and data reduction}
\label{sec:observations}
\subsection{LOFAR HBA data}
\label{sec:hba}
Data from the LOFAR High-Band Antenna (HBA) system were acquired during Cycle~0
observations in February 2013 (see
Table~\ref{tab:observations} for details). We used the
HBA\verb|_|Dual\verb|_|Inner configuration resulting in a field of view (FOV) of approximately
$8\degr$ with baseline lengths of up to 85~km. We used the 8-bit
mode that can provide up to 95~MHz of instantaneous
bandwidth.\footnote{The data sent from the LOFAR stations is encoded with 8-bit
  integers.} The data were first
processed using the `demixing' technique \citep{van_der_tol_07a} to remove
Cas~A and Cyg~A from the visibilities. In order to reduce the data
storage volume and speed up the processing, the data
were then averaged in time, from 1 to 10~s, and in frequency, from 64 channels to
1 channel per sub-band, prior to further data reduction.\footnote{The software used for the LOFAR data reduction is documented in
  the `LOFAR Imaging Cookbook' available at
  \href{https://www.astron.nl/radio-observatory/lofar/lofar-imaging-cookbook}{https://www.astron.nl/radio-observatory/lofar/lofar-imaging-cookbook}} We
note that the strong compression in time and frequency will lead to bandwidth
smearing in the outer parts of the FOV; since we are only interested in the
central $1\degr$, this does not affect our science goals.

Before calibration, we applied the New Default Processing Pipeline
({\small NDPPP}) in order to mitigate the radio frequency interference (RFI)
using {\small AOFlagger} \citep{offringa_10a}. Following this, we calibrated phases and amplitudes of our
primary calibrator using Black Board Selfcal \citep[{\small
  BBS};][]{pandey_09a}, where we used the calibrator model and flux density of
3C~48 given by
\citet{scaife_12a}. The gain solutions were transferred to the target and sets
of 18
sub-bands were combined into 18 bands each with a bandwidth of $3.5~\rm MHz$, before we performed a calibration in
phase only, employing the global sky model (GSM) developed by \citet{scheers_11a}.

For self-calibration, we imaged
the entire FOV with the {\small AWIMAGER} \citep[][]{tasse_13a}. We
used a {\small CLEAN} mask, created with the Python Blob Detection
and Source Measurement \citep[{\small
  PyBDSM};][]{mohan_15a} software. We used a sliding window of 420~arcsec
to calculate the local rms noise, which suppresses sidelobes or
artefacts 
in the mask and thus in the {\small CLEAN} components. A sky model was then created with {\small
CASAPY2BBS.PY}, which converts the model image ({\small CLEAN} components) into a {\small
BBS}-compatible sky model. We determined direction independent phase
solutions with a 10~s solution interval and applied these solutions with {\small
  BBS}. In order to test the success of our calibration, we checked the peak
and integrated flux density of 3C~34, a compact ($<$$1~\rm arcmin$) bright (17~Jy at
150~MHz) source $0\fdg 9$
south-east of 3C~31. The first round of self-calibration in phase increased the peak flux density
of 3C~34 by 8~per cent, but did not change the integrated flux density of
either 3C~31 or 34 by more than 1 per cent. 

For a second round of
phase-only self-calibration, we imaged the FOV
using the Common Astronomy Software Applications \citep[{\small CASA};][]{mcmullin_07a}
and used the MS--MFS {\small CLEAN} algorithm described by \citet{rau_11a}. A sky model was
created with {\small PyBDSM}. We subtracted all sources in the FOV, except
3C~31, our target, from the $(u,v)$ data and the sky model. Following this, we performed another direction-independent calibration
in phase with {\small BBS}. This resulted in a significant improvement of our
image: the peak flux densities of components within 3C~31 increased by up to 40 per cent, and
the rms noise decreased by 20--30 per cent, while the integrated flux
densities were preserved. This self-calibration also suppressed some `striping' surrounding 3C~31, which was
obvious in the map before this last round of self-calibration. This method is a simplified way of performing a `facet
  calibration' \citep{van_weeren_16a,williams_16a}, where we use only one facet, namely
  our source 3C~31. The solutions, while technically using {\small BBS}'s
  direction-independent phase solutions, are tailored for the direction of
  3C~31, resulting in an improved image. Our rms noise is still a factor of
  3--5 higher than the expected thermal noise level ($\approx$$100~\umu\rm
  Jy\,beam^{-1}$). This is partly due to the imperfect subtraction of the
  field sources, which leaves behind residual sidelobes stemming from the variations
  in phase and amplitude due to the ionosphere. In order to reach the thermal
  noise level, approximately ten of the brightest field sources would have to be subtracted with
  tailored gain solutions (D.~A.~Rafferty 2017, priv.\ comm.). But since our 145-MHz map is our
  most sensitive one, showing the largest angular extent of 3C~31, and we are
  interested in measuring the spectral index with the maps at other
  frequencies, it is not necessary for this work to reach the thermal noise
  limit and we can achieve our main science goals with the map created with
  our simplified method.

We created the final image using {\small CASA's} MS--MFS {\small CLEAN}, fitting for the
radio spectral index of the sky model (`nterms=2') and using Briggs' robust weighting (`robust=0')
as well as using angular
{\small CLEAN}ing scales of up to 800~arcsec. The rms noise level
of the map depends only weakly on position: it is largely between $0.4$ and
$0.5$~$\rm mJy\,beam^{-1}$. The map has a resolution of $16.5\times 11.8~\rm arcsec^2$
($\rm PA=73\degr$).\footnote{Angular resolutions in this paper are referred to as the full width at half maximum (FWHM).} The theoretical resolution is in good agreement with
the actual resolution which we measured by fitting 2D Gaussians with {\small IMFIT} (part of the Astronomical Image Processing System; {\small AIPS}) to point-like sources.\footnote{{\scriptsize AIPS} is free software available from the NRAO.} We did not correct for primary beam attenuation: the
correction is smaller than 3 per cent across our target. We checked that the
in-band spectral indices agree with the overall (across all frequencies)
spectral index. It is a recognized issue that the overall LOFAR flux scale, and the in-band HBA spectral index
can be significantly in error after the standard gain transfer process due to the uncertainty in the `station efficiency factor'
as function of elevation \citep{hardcastle_16b}. The fact that this is not
an issue for us is probably related to the fact that the primary calibrator is
very nearby ($6^\circ$ away) and at a similar declination, so that the difference in elevation does not play a
role.

\subsection{LOFAR LBA data}
Data with the LOFAR Low-Band Antenna (LBA) system were acquired during Cycle~0
observations in 2013 March. The LBA
observational setup differs slightly from that of HBA: we observed with two beams
simultaneously, one centred on our target, 3C~31, and the other centred on
our calibrator, 3C~48 (see
Table~\ref{tab:observations} for details). After removing RFI with the {\small AOFlagger}, the data were demixed to remove Cas~A and Cyg~A from the visibilities. We averaged from 1 to 10~s
time resolution and from 64 to 4 channels per sub-band. Time-dependent phase and
amplitude solutions for 3C~48 were determined using the source model and flux
scale of \citet{scaife_12a} and transferred to our
target. Following this, the target data
were again calibrated in phase only with {\small BBS} using the GSM sky model. For
self-calibration in phase, we imaged the FOV with {\small AWIMAGER},
using the same mask as for the HBA data and converted the {\small CLEAN} components
into a {\small BBS}-compatible sky model with {\small CASAPY2BBS.PY}. We calibrated in phase with {\small BBS} using
a solution interval of 10~s and corrected the data accordingly. After
  self-calibration the peak
  flux density of 3C~34 increased by 8 per cent; the integrated flux density
  of 3C~31 increased by 3~per cent, probably due to a slightly improved
  deconvolution. We attempted a
second round of phase calibration (as with the HBA data) with everything but 3C~31 subtracted from
the $(u,v)$ data (and the sky model), but this did not result in any further
improvement and reduced the integrated flux density of 3C~31 by 5~per
cent. We therefore used the maps with one phase calibration only in the remainder of our analysis.
\begin{figure*}
  \includegraphics[width=1.0\hsize]{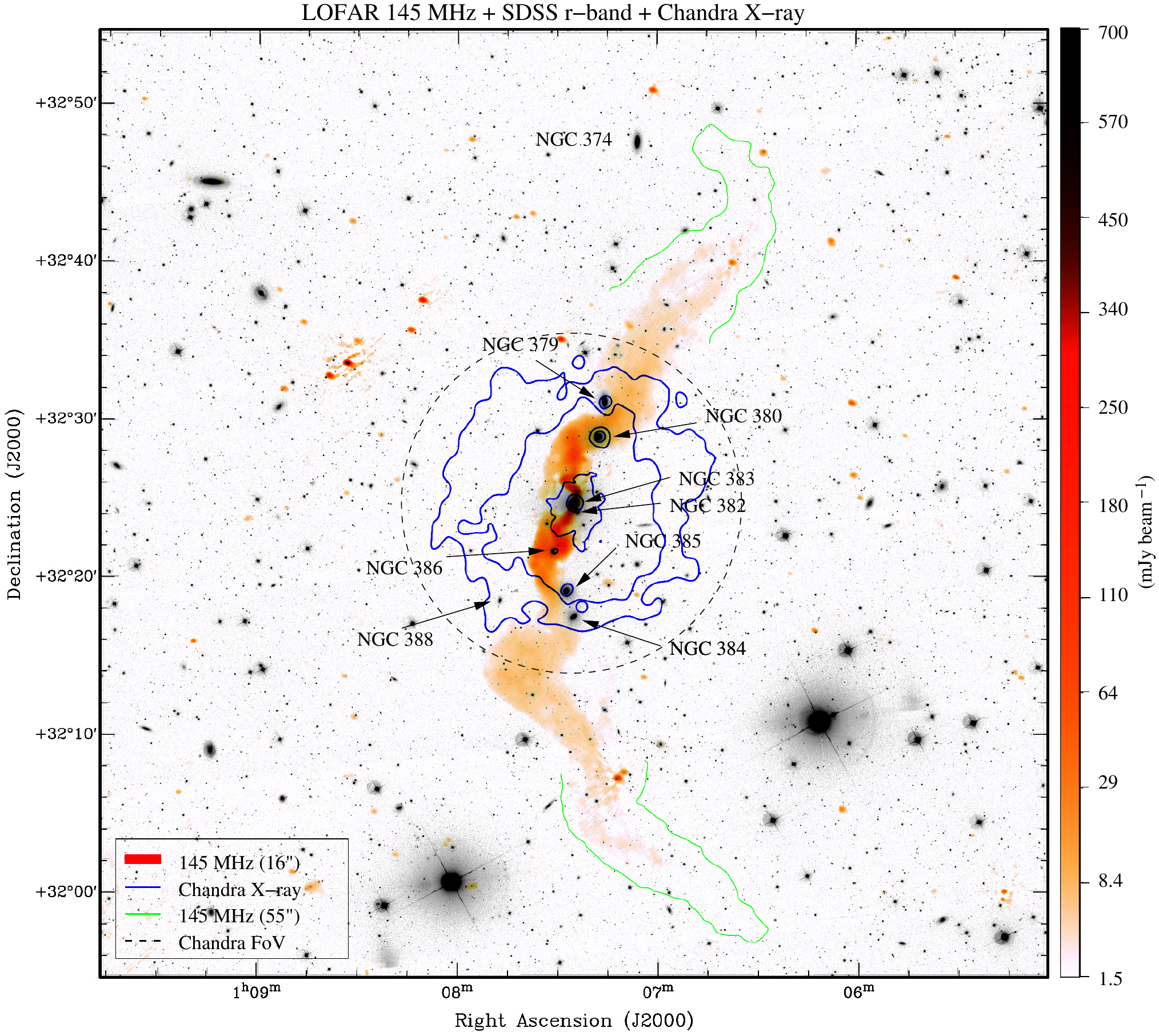}
    \caption{LOFAR 145-MHz radio continuum emission, overlaid on a SDSS
      r-band map with \emph{Chandra} X-ray emission as contours. The image
      shows a $1~\rm deg^2$ area (corresponding to $1.2\times 1.2$~Mpc$^2$), centred on
      3C~31, where the 145-MHz radio continuum emission is shown with a red colour
      transfer function ($1.5\ldots 700~\rm mJy\,beam^{-1}$) at $16.5\times
      11.8~\rm arcsec^2$ ($\rm PA=73\degr$)
    resolution. The blue contours show the hot gas of the ICM as
    traced by \emph{Chandra} X-ray emission (contours at $0.7$, $0.84$, $1.2$
    and $2.4$ ACIS counts\,s$^{-1}$), where the dashed circle indicates the FOV of the
    \emph{Chandra} observations. Green contours indicate the maximum extent of
    the faint radio tails as measured from our low-resolution map (contours
    are at $4.5~\rm mJy\, beam^{-1}$).}
  \label{fig:HBA}
\end{figure*}

As for the HBA observations, we performed the final imaging of the data within {\small CASA}, using the MS--MFS {\small CLEAN}
algorithm with the spectral index fitting of the sky model enabled (`nterms=2') and a variety of
angular scales. The rms noise of the LBA image
is 5~$\rm mJy\,beam^{-1}$ using robust weighting (`robust=0') at a resolution of $38.0\times
23.5~\rm arcsec^2$ ($\rm PA=-73\degr$). The theoretical resolution is in good
agreement (within 5~per cent) of the actual resolution, as for the LOFAR HBA
data (Section~\ref{sec:hba}).

\subsection{VLA data}
Observations with the Karl G.\ Jansky Very Large Array (VLA) with the
recently commissioned $P$-band receiver were taken between 2013
December and 2014 December (see Table~\ref{tab:observations} for details). We followed standard data
reduction procedures, using {\small CASA} and utilizing
the flux scale by \citet{scaife_12a}. We checked that no inverted
cross-polarizations were in the data (M.\ Mao \& S.~G.\ Neff 2014, priv.\
comm.). Of the 16 spectral windows 5 had to be discarded (1, 2, 11, 14 and
15) because the primary calibrator 3C~48 had no coherent
phases, which prevents calibration. A further three spectral windows (0, 8 and 9)
had to be flagged because of strong RFI. We imaged the data in {\small CASA}, first in
A-array alone to self-calibrate phases, then in A- and B-array, and finally in
A-, B- and C-array together. Prior to combination we had to change the
polarization designation from circular polarizations (Stokes $RR$, $LL$) to linear
polarization (Stokes $XX$, $YY$) of the A- and B-array data.\footnote{The VLA $P$-band feeds are linearly polarized, but the A- and B-array observations were in error labelled as circular which would have prevented a combination with the C-array data.} Furthermore,
we had to re-calculate the weights of the C-array data which were unusually
high ($\approx$$10000$). We performed two rounds of self-calibration in
phase only followed by two rounds of self-calibration in phase and
amplitude with solution intervals of 200~s. We normalized the amplitude gains and found that the
integrated flux densities of 3C~31 and 34 did not change by more than 2 per
cent. The rms noise of the full bandwidth VLA $P$-band image is
$0.15~\rm mJy\,beam^{-1}$ at a resolution of $7.5\times 4.6~\rm arcsec^2$ ($\rm
PA=71\degr$). The largest angular scale the VLA can detect at 360~MHz is
$1\fdg 1$ in C-array, which is enough to image 3C~31 to the same angular extent
as measured with LOFAR.

\begin{figure*}
  \includegraphics[width=1.0\hsize]{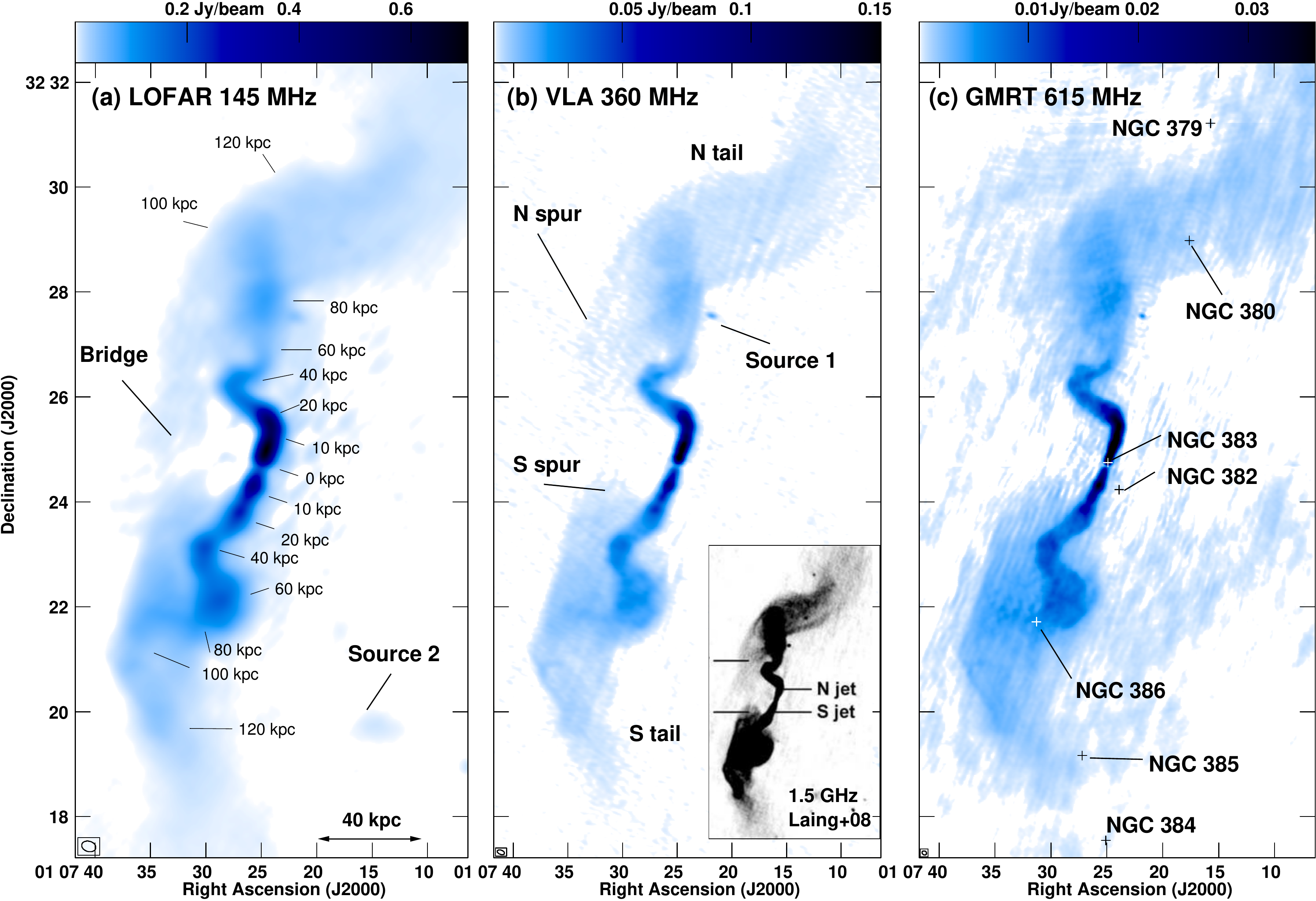}
  \caption{\emph{Left:} Zoom-in on the combined LOFAR HBA data at
    115--178~MHz, with a mean frequency of 
    145~MHz, at an angular resolution of $16.5\times 11.8~\rm arcsec^2$
    ($\rm PA=73\degr$). The map shows the central 14~arcmin (290~kpc), centred on the
    nucleus. The transfer function ($1.5$--700~mJy\,beam$^{-1}$) is stretched
    to show weak, diffuse emission. The rms noise is approximately
    $0.5$~mJy\,beam$^{-1}$. \emph{Middle:} Combined VLA data at 260--420~MHz
  (mean 360~MHz) at an angular resolution of
  $8.7\times 4.7~\rm arcsec^2$ ($\rm PA=71\degr$). The grey-scale
  corresponds to intensities of $0.05$--150~mJy\,beam$^{-1}$. The rms noise level is
  $0.25~\rm mJy\,beam^{-1}$. The inset shows the $1.5$-GHz map of
  \citet{laing_08a} at $5.5$~arcsec resolution. \emph {Right:} Combined GMRT data at 615~MHz at an angular
  resolution of $5.4\times 4.5~\rm arcsec^2$ ($\rm PA=56\degr$). The grey-scale
  corresponds to intensities of $0.15$--36~mJy\,beam$^{-1}$. The rms noise
  level is $0.3~\rm mJy\,beam^{-1}$. In panels (a)--(c), we label features
  discussed in the text with the size of the synthesized beam shown in the
  bottom-left corner. Panel (c) shows the position of the cluster
  galaxies.}
\label{fig:3c31_maps_zoom}
\end{figure*}

\subsection{GMRT data}
Observations with the Giant Metrewave Radio Telescope (GMRT) were taken in August
2007 in 615/235~MHz dual band mode \citep[][see Table~\ref{tab:observations} for details]{swarup_91a}. In this mode, the
frequencies are observed simultaneously in two 16-MHz sub-bands consisting of
$128\times125$-kHz channels. The 615-MHz data are stored in one polarization ($RR$) in
both the upper and lower sideband. The data reduction was carried out following standard flagging and
calibration
 procedures in {\small AIPS}. The FWHM of
the GMRT primary beam at 615~MHz is 44~arcmin; we corrected for primary beam attenuation
in our final images with {\small PBCOR} (part of {\small AIPS}). In order
to accurately recover the flux and structure of 3C~31, we imaged all bright
sources well beyond the primary beam, using faceting to account for the sky
curvature.  The extent of 3C~31 is such that at 615~MHz, two facets were
necessary to cover the entire source, one for each radio tail. The seam
between the two facets was placed such that it crosses between the bases of the
two jets where no radio emission is observed. The facets were iteratively
{\small CLEAN}ed to reveal increasingly faint emission and then 
re-gridded on the same coordinate system and linearly combined with {\small FLATN}  (part of {\small AIPS}) to create the
final image. The GMRT 615-MHz map has a final synthesized beam of 
$5.4\times4.5~\rm arcsec^2$ ($\rm PA=56\degr$) and the rms noise is $0.3$~mJy\,beam$^{-1}$.

\subsection{General map properties}
In addition to the new reductions described above, we used a $609$-MHz map observed with the Westerbork
Radio Synthesis Telescope (WSRT), which we obtained from the `Atlas of
DRAGNs' public webpage which presents 85 objects of the `3CRR' sample
of \citet{laing_83a}.\footnote{`An Atlas of DRAGNs', edited by J.~P.\
  Leahy, A.~H.\ Bridle and R.~G. Strom,
  \href{http://www.jb.man.ac.uk/atlas}{http://www.jb.man.ac.uk/atlas}} This map, which was presented
by \citet{strom_83a}, has an angular resolution of 55~arcsec, the
lowest of our 3C~31 maps. The WSRT 609-MHz map has the advantage over
the GMRT 615-MHz map that it recovers better the large-scale structure
of 3C~31, which is largely resolved out in the high-resolution GMRT
map. Hence, for the spectral analysis in what follows we use only the WSRT 609-MHz map
while the GMRT 615-MHz map is used only for the morphological
analysis. The maps created from the LOFAR and VLA observations,
together with the WSRT map, will be used in the rest of the paper to
study the spatially resolved radio spectral index of 3C~31.

In order
to ensure that we have a set of maps that are sensitive to the same angular scales, we imaged all our data 
with an identical $(u,v)$-range between $0.04$ and
$4.9$~k$\lambda$. Since we did not have the the $(u,v)$ data for the 609-MHz map available, we could not re-image these data. But the $(u,v)$-range for these data is with $0.06\rightarrow 5.6~{\rm k}\lambda$ quite similar, so re-imaging is not necessary. In what follows, these maps are referred to as `low-resolution maps', whereas the maps without a $(u,v)$-range applied are
referred to as `full-resolution maps'.

For further processing, we took the maps into {\small AIPS}, where we
made use of {\small PYTHON} scripting interface {\small PARSELTONGUE}
\citep{kettenis_06a} to batch process them. We used {\small
  CONVL} to convolve them with a Gaussian to the same angular resolution and
{\small HGEOM} to register them to the same coordinate system. To create a
mask, we blanked unresolved
background sources with {\small BLANK}, and applied the same mask to all maps
to be compared. The flux densities as determined from the final maps are in agreement with
literature values. The integrated spectrum of 3C~31 can be described by
a power law between $30$ and $10700$~MHz with a radio spectral index of $-0.67\pm 0.01$. We use calibration uncertainties of 5
per cent and add noise contributions from the rms map noise and zero-level
offset in order to determine our uncertainties (see Appendix~\ref{flux} for details).

\begin{figure*}
  \includegraphics[width=1.0\hsize]{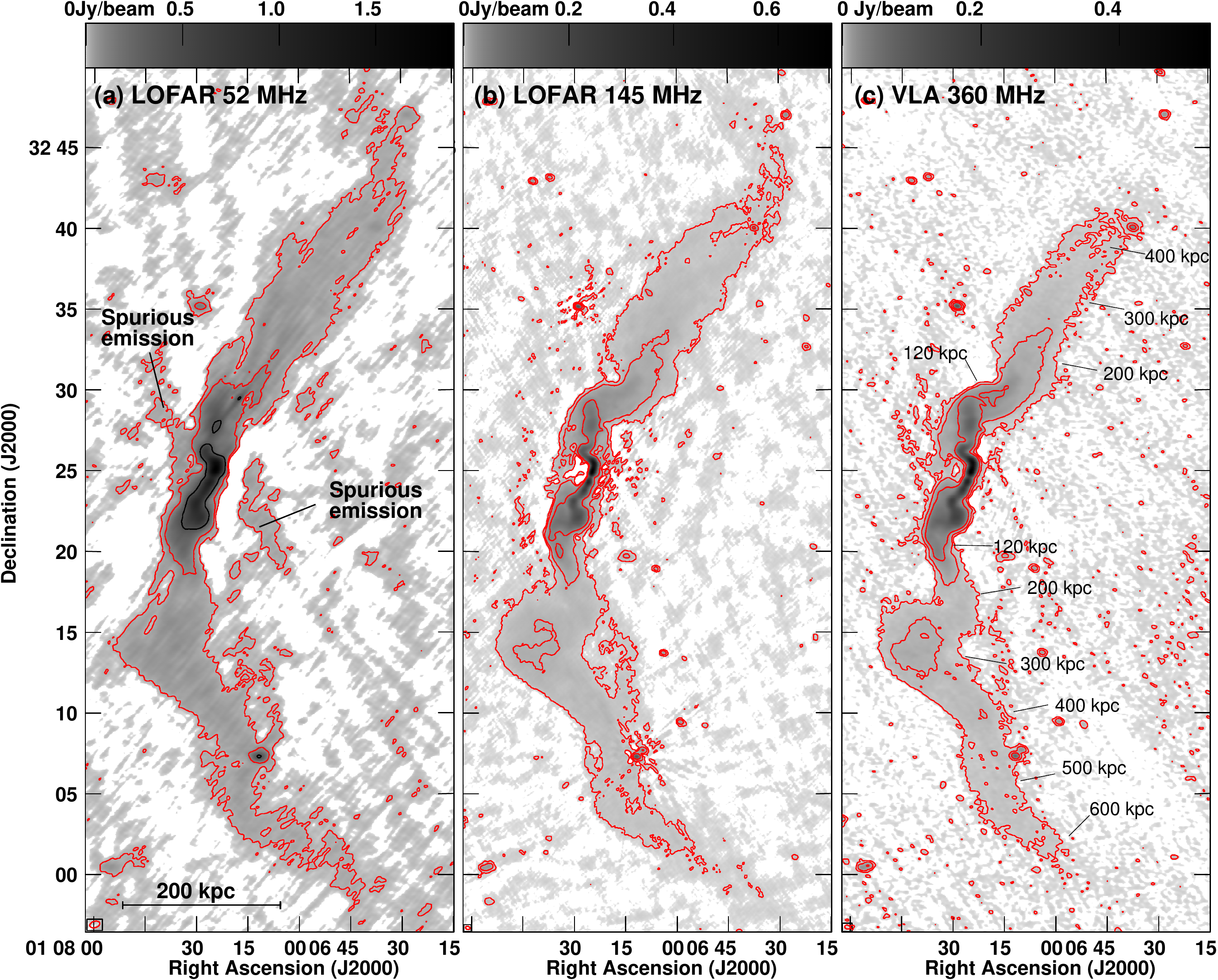}
  \caption{\emph{Left:} Combined LOFAR LBA data at 35--85~MHz (mean 52~MHz),
    at an angular resolution of $38.0\times 23.5~\rm arcsec^2$ ($\rm PA=-73\degr$).  The transfer function (0--1900~mJy\,beam$^{-1}$) is stretched
    to show weak, diffuse emission. The rms noise is
  $6.5$~mJy\,beam$^{-1}$. \emph{Middle:} Combined LOFAR HBA data at 115--178~MHz (mean 145~MHz), at an angular resolution of $16.5\times 11.8~\rm arcsec^2$
    ($\rm PA=73\degr$). The grey-scale
  corresponds to intensities of 0--700~mJy\,beam$^{-1}$. The rms noise is $0.5~\rm mJy\,beam^{-1}$. \emph{Right:} Combined VLA $P$-band data at 288--432~MHz (mean 360~MHz), at an angular
    resolution of $21.0\times 14.6~\rm arcsec^2$ ($\rm PA=74\degr$). The grey-scale
  corresponds to intensities of 0--530~mJy\,beam$^{-1}$. The rms noise level is $0.3~\rm
    mJy\,beam^{-1}$. In all panels, contours are at $2.5$, $12.5$ and $62.5$
    $\times$ the rms noise level and the size of the synthesized beam is shown in the
  bottom-left corner.}
  \label{fig:3c31_maps}
\end{figure*}
\begin{figure*}
  \includegraphics[width=1.0\hsize]{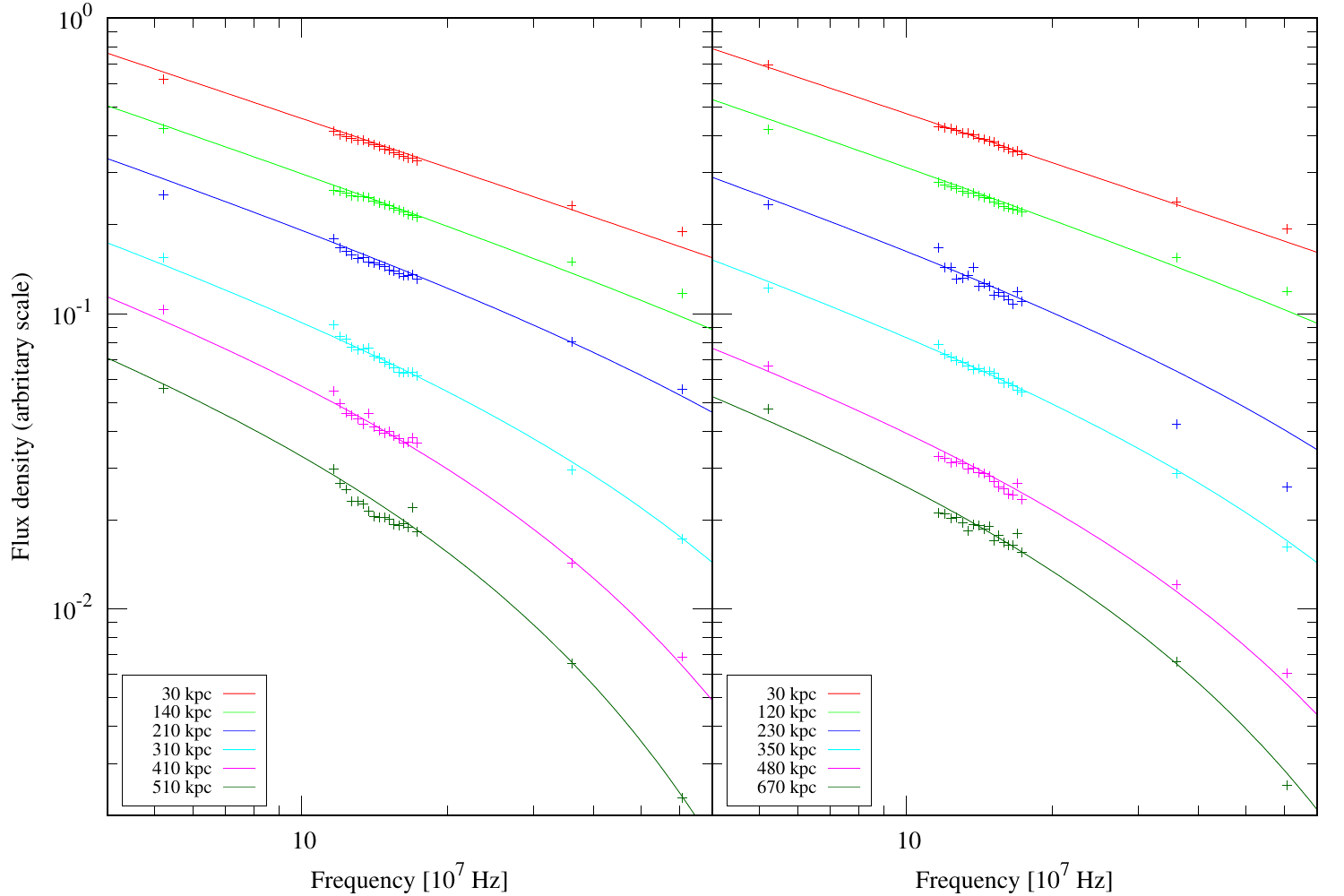}
  \caption{Radio continuum spectra within selected regions between 52 and 609
    MHz at 20 frequencies (LOFAR LBA 52~MHz, 17 combined sub-bands at
    117--173~MHz from LOFAR HBA, VLA 360~MHz and WSRT 609~MHz). Solid lines are the spectra as predicted by our best-fitting advection models. The size of the data
    points is approximately equivalent to the size of the error bars. Spectra
    in the northern tail are shown in the \emph{left} panel, spectra in the
    southern tail are shown in the \emph{right} panel. The legend shows the
    distance to the nucleus (see Figs.~\ref{fig:3c31_maps_zoom}c and \ref{fig:3c31_maps}c).}
\label{fig:age_spectrum}
\end{figure*}

\section{Morphology and observed spectrum}
\label{sec:morphology_and_observed_spectrum}
\subsection{Morphology}
\label{morphology}
Figure~\ref{fig:HBA} presents a panoramic low-frequency view of 3C~31, showing
our 145-MHz map at full resolution as colour-scale, with contours
of \emph{Chandra} X-ray emission, showing the hot ICM. We overlay the data on a SDSS r-band image, which shows the optical
stellar light. The
SDSS r-band image shows the large, diffuse halo of the host galaxy, NGC~383, with a diameter of
3~arcmin (60~kpc). The closest 
member of the group is NGC~382, $0.5$~arcmin (10~kpc)
south-south-west of NGC~383. There are a handful of further members of this
group, some of which can be detected
in the X-ray data as point-like sources embedded in the diffuse X-ray emission
from the ICM. None shows any radio emission in our image.

Figure~\ref{fig:3c31_maps_zoom} shows the full-resolution maps of
the central 14~arcmin (290~kpc) at 145 (LOFAR HBA), 360 (VLA) and 615~MHz (GMRT). Our 360-MHz map shows the northern and southern `spurs' of
\citet{laing_08a} -- extensions from the tails towards
the nucleus. Both spurs are also visible
in the 145-MHz map, where the northern spur is connected via a `bridge'
to the southern one. Similarly, the 615-MHz map shows both spurs although the
visibility of the northern one is limited by the image fidelity. We find two compact sources in the radio
continuum near the jet (labelled as `Source 1' and `Source 2' in Fig.~\ref{fig:3c31_maps_zoom}), but they have no
counterparts in the optical (SDSS), the mid-infrared \citep[\emph{WISE} 22~$\umu$m;][]{wright_10a} or
in the far-UV (\emph{GALEX}). We thus conclude that these sources are unrelated to 3C~31 or the galaxies associated with it.

Further away from the nucleus, the radio tails can be particularly
well traced at low frequencies. Figure~\ref{fig:3c31_maps} presents
our maps at full resolution (although we omit the A-array from the VLA
data) at 52 (LOFAR LBA), 145 (LOFAR HBA) and 360~MHz (VLA); the
emission extends spatially further than any of the previous studies
have revealed.  \citet{andernach_92a} found an angular extent of
40~arcmin in declination in their 408-MHz map and speculated that the
tails did not extend any further. Our new data show that 3C~31 extends
at least 51~arcmin mostly in the north-south direction, between
declinations of $+31\degr 57\arcmin$ and $+32\degr 48\arcmin$, as
traced by our most sensitive 145-MHz map. The projected linear size
hence exceeds 1~Mpc, so that 3C~31 can now be considered to be a
member of the sub-class of objects that are referred to in the
literature as giant radio galaxies (GRGs). It remains difficult to
ascertain whether we have detected the full extent of 3C~31 since it is certainly possible that the radio emission extends even further, as 3C~31
has tails of diffuse emission extending away from the nucleus -- as
opposed to lobes with well-defined outer edges \citep[see][for a LOFAR
study of Virgo~A, exhibiting well-defined lobes]{deGasperin_12a}.
 
 \begin{figure*}
  \includegraphics[width=1.0\hsize]{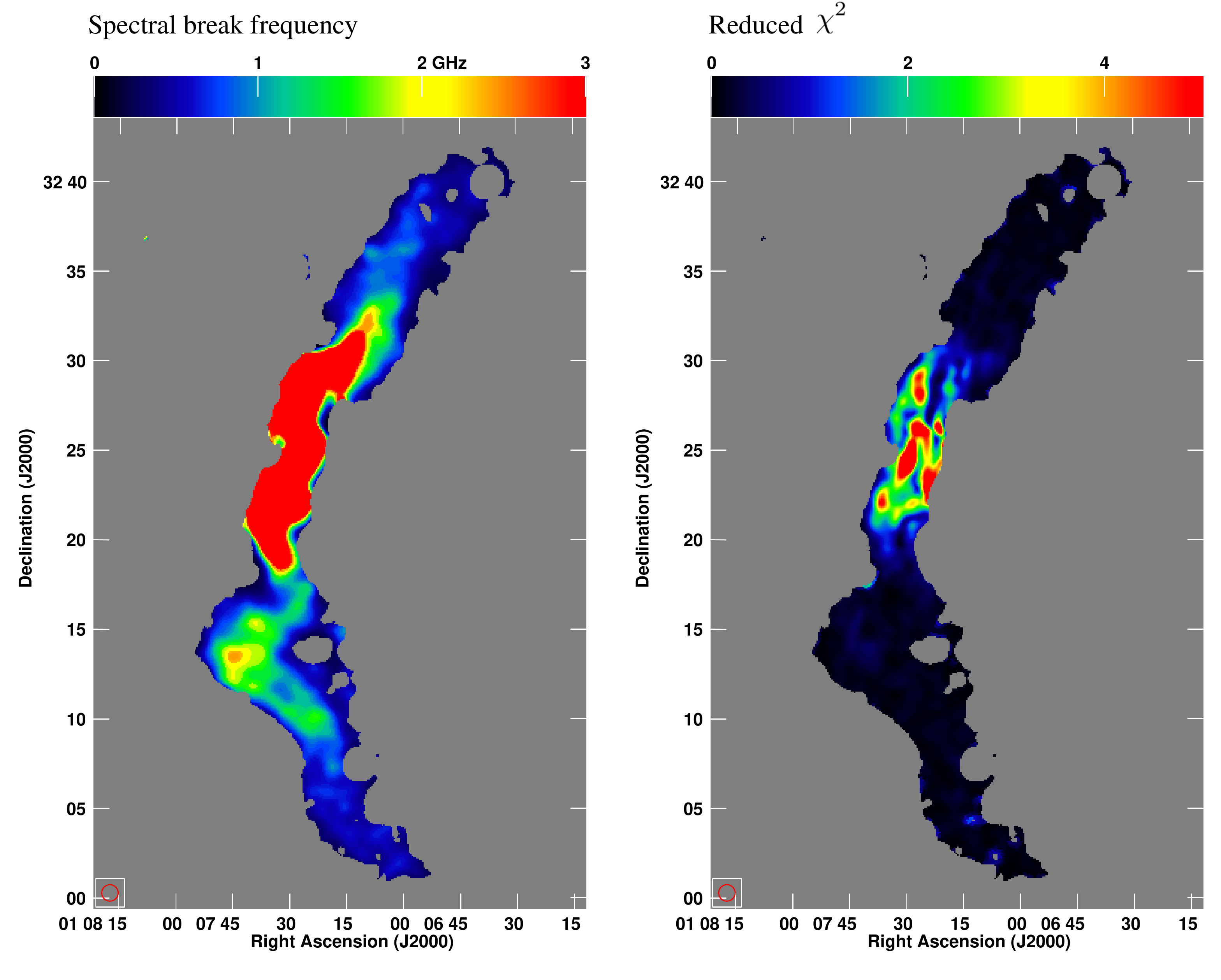}
  \caption{\emph{Left:} Spectral break frequency at 55~arcsec
    angular resolution (equivalent to 19~kpc) as indicated by the boxed circle
    in the bottom-left corner. Data points at 20 frequencies 
    (LOFAR LBA 52~MHz, 17 combined sub-bands at
    117--173~MHz from LOFAR HBA, VLA 360~MHz and WSRT 609~MHz) were fitted
    with a JP model (assuming $\delta_{\rm inj}=2.1$). \emph{Right:} The
    corresponding reduced $\chi^2$.}
\label{fig:break}
\end{figure*}
\citet*{wezgoviec_16a} claimed the detection of diffuse $1.4$-GHz emission surrounding 3C~31, based on combining NVSS and Effelsberg single-dish imaging data. Their integrated flux density of 3C~31 lies $\approx$$1.2$~Jy (or $3.6~\sigma$) above our integrated power-law spectrum (see Appendix A), which predicts a $1.4$-GHz flux density of $5.61$~Jy. We find no hint of this `halo' structure. The largest detectable angular scale in our HBA map is $\approx$$1\fdg 4$ (a factor $1.6$ larger than the detected total source extent), so that we should not be resolving out cocoon structure on scales comparable to the detected source. Our 145-MHz low-resolution map has a $3\sigma$ threshold in surface brightness of $\approx$$1.3~\umu$Jy arcsec$^{-2}$. If we consider the 1.4-GHz extended halo feature reported by \citet{wezgoviec_16a} and assume a halo flux of $\approx$$1.2$ Jy (the excess emission above our integrated spectrum) distributed uniformly over a surface area of $\approx$$227$ arcmin$^2$, then for $\alpha$ between $-0.7$ and $-1.5$, the 145-MHz surface brightness would be between $7$--$45~\umu$Jy arcsec$^{-2}$. This is well above our detection threshold, and even bearing in mind the likely uncertainty in the flux estimate from \citet{wezgoviec_16a}, our data are not consistent with the presence of the large-scale halo presented in that work. However, we cannot rule out the existence of an even larger, lower surface brightness halo, which could still evade detection in our observations. This possibility is discussed in Section~\ref{source_age}.

\subsection{Spectrum}
\label{sec:spectrum}
In Fig.~\ref{fig:age_spectrum}, we show the evolution of the radio continuum
spectrum as function of distance from the nucleus. The spectrum is a power law
within 120--140~kpc from the nucleus, so that spectral ageing plays no
role there, at least not in the frequency range covered by our data. At larger distances,
a significant spectral curvature develops, which grows with increasing
distance from the nucleus. The spectra can be characterized by the fitting of Jaffe--Perola \citep[JP;][]{jaffe_73a} models to determine a characteristic (`break') frequency
\citep[e.g.][]{hughes_91a}. We fitted models to our low-resolution maps using the Broadband Radio Astronomy Tools \citep[{\small
  BRATS};][]{harwood_13a}, assuming a CRE injection spectral index of $\delta_{\rm inj}=2.1$ (with a CRE number density of $n\propto E^{-\delta_{\rm inj}}$, where $E$ is the CRE energy). The resulting spectral break frequency is shown in
Fig.~\ref{fig:break}. The high reduced $\chi^2$ found in the inner region can
be at least in part explained due to averaging over regions with very
different spectra \citep[cf. fig.~12 in][]{laing_08a}. We used for the fitting 17 combined sub-bands from LOFAR HBA, but we tested that using only one HBA map instead changes the spectral age locally by 20 per cent at most and averaged across the tails the difference is even smaller.

As expected, the spectral break frequency decreases with increasing distance from the nucleus, which can be explained by spectral ageing. The spectral break frequency 
$\nu_{\rm brk}$ can be related to the age $\tau_{\rm Myr}$ of the CREs (in units of Myr) via \citep[e.g.][]{hughes_91a}:
\begin{equation}
  \nu_{\rm brk} = 2.52 \times 10^3 \frac{B/{10~\umu\rm G}} {[(B/{10~\umu \rm G})^2 + (B_{\rm CMB}/{10~\umu \rm G})^2]^2\tau_{\rm Myr}^2}~{\rm GHz},
\label{eq:nu_brk}
\end{equation}
where $B$ is the magnetic field strength
and $B_{\rm CMB}$, the equivalent cosmic microwave background (CMB) magnetic field
strength of $3.2~\umu\rm G$ (at redshift zero), is
defined so that the magnetic energy density is equal to the CMB photon
energy density. If we use a magnetic field strength of $B=5~\umu\rm G$,
which is an extrapolation from \citetalias{croston_14a}, we find that for a spectral break
frequency of 1~GHz, the CREs are 100~Myr old. This already provides a
good estimate of the source age. However, the magnetic field
strength (whether estimated via equipartition or otherwise) is
expected to change with distance from the nucleus, and the average
field strength may be different to the value we assume here. In the next section, we introduce a model of cosmic-ray transport in order to
investigate the evolution of tail properties in more detail.

\section{Cosmic-ray transport}
\label{sec:cr_transport}

\begin{figure*}
  \includegraphics[width=1.0\hsize]{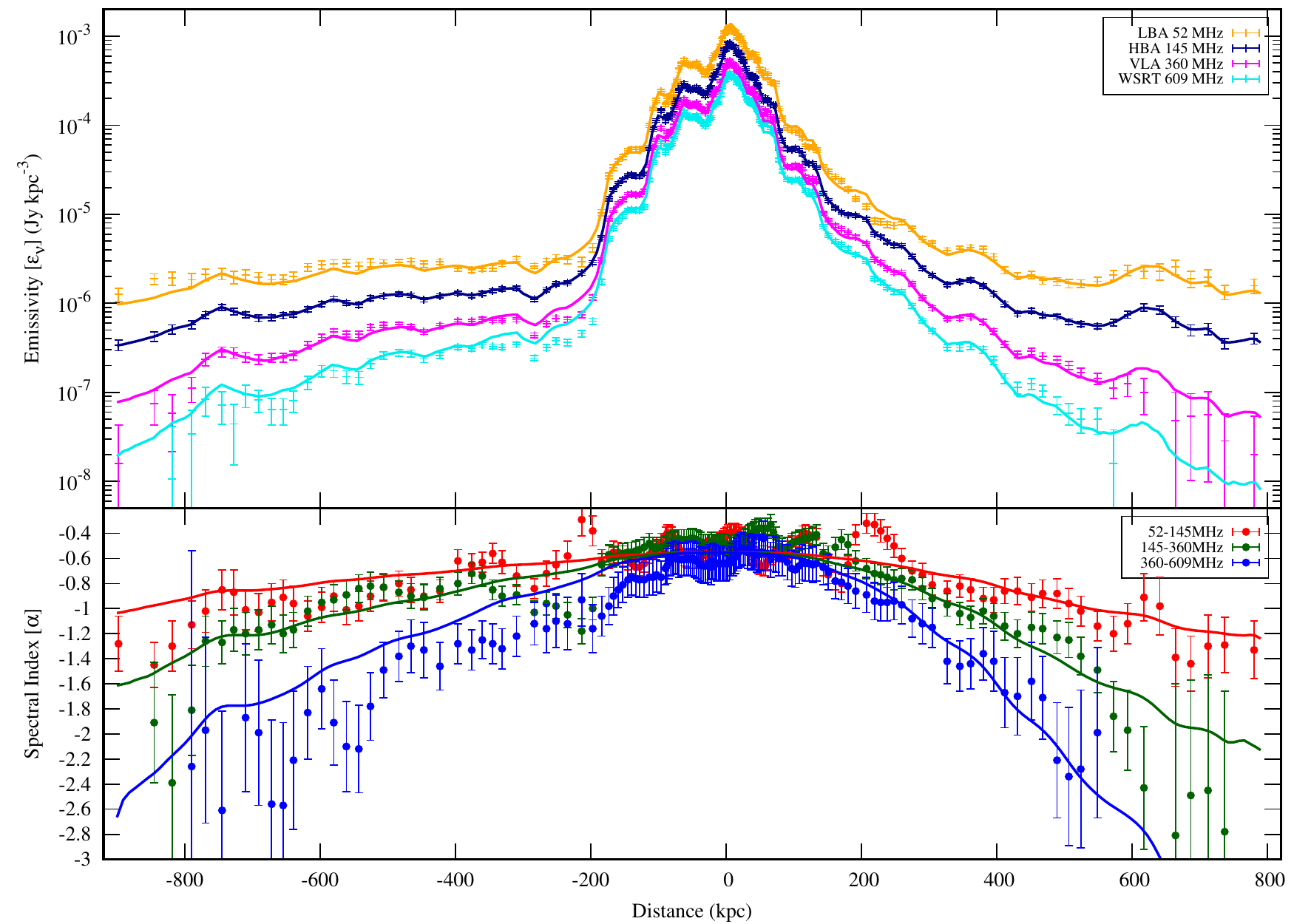}
  \caption{\emph{Top panels:} Profiles of the radio continuum emissivity as function of
    distance from the nucleus at various frequencies. Shown are the LOFAR LBA 52~MHz (orange), HBA 145~MHz
    (dark blue), VLA 360~MHz (magenta) and WSRT 609~MHz (cyan) observations.  \emph{Bottom panels:} Corresponding
    profiles of the radio spectral index between 52 and 145~MHz (red), 145
    and 360 MHz (dark green) and between 360 and 609~MHz (blue). Solid lines
  show the best-fitting advection models using {\scriptsize
    SPINNAKER} (see text for details). The northern tail is on the right-hand side ($z>0$~kpc) and the southern tail is on the left-hand side ($z<0$~kpc).}
\label{fig:alpha}
\end{figure*}

To explore the tail physical conditions and dynamics in more detail we
use the quasi-1D cosmic-ray transport model of
\mbox{\citet[][hereafter \citetalias{heesen_16a}]{heesen_16a}}. In that
work, cosmic-ray transport via advection and diffusion is considered within
galaxy haloes. For a jet environment we expect advection to be the
dominant transport mode. This is corroborated by the intensity
profiles, which are of approximately exponential shape as expected for
an advection model (\citetalias{heesen_16a}) and the corresponding
radio spectral index profiles, which show a linear steepening
(Fig.~\ref{fig:alpha}).

The model self-consistently calculates the evolution of the
CRE spectrum along the tails, accounting for
adiabatic, synchrotron and inverse Compton (IC) losses, with the aim of
reproducing the evolution of radio intensity and spectral index, as
shown in Fig.~\ref{fig:alpha}. The aim is to obtain a self-consistent
model for the velocity distribution and the evolution of magnetic field
strength along the tails.  We make the following basic assumptions:
 (i) the tail is a steady-state flow, (ii) the CRE energy distribution is described by a power law at 
the inner boundary, and (iii) there is no in-situ particle acceleration
in the modelled region. The first assumption is not strictly valid in the outermost parts of the source, as we expect that the tails are still growing; we discuss the effects of expansion of the outer tails in Section~\ref{sec:results}. The second assumption is well justified by the
 radio spectral index in the inner jet (Fig.~\ref{fig:age_spectrum}). It
is difficult to obtain any direct constraints on particle acceleration
in the outer parts of FR~I plumes, and so the assumption of no particle
acceleration is a limitation of all spectral ageing models. 

We model the tails between 15 kpc and 800 (northern) or 900 (southern tail) kpc, taking 15~kpc as the inner boundary, as the jet parameters at this distance are
well constrained from the work of \citet{laing_02a,laing_02b}. The
geometry of the tails is taken from observations assuming a constant
inclination angle of $52\degr$ to the line of sight
\citep{laing_02a}. The details of the advection modelling process are
described in Section~\ref{sec:advection_model}. There are a number of free
parameters in the model, and so in the next section we summarize our
approach to exploring the parameter space of possible models.

\subsection{Modelling approach and assumptions}
\label{sec:modelling_approach_and_assumptions}

The model parameters are as follows, where the injection parameters are at 15~kpc distance from the nucleus:
\begin{itemize}
\item Injection magnetic field, $B_{0}$
\item Injection velocity, $\varv_{0}$
\item The deceleration parameter, $\beta$
\end{itemize}
with the primary model outputs being a profile of velocity and
magnetic field strength along the tail. The expansion of the tails leads to strong
adiabatic losses along the tail which are inconsistent with the
observed spectral evolution unless the flow is decelerating. We chose
to model such a deceleration assuming $\varv \propto r^{\beta}$, where $r$
is the tail radius. We also tested a velocity dependence on distance
rather than tail radius; however, this led to poorer model fits. The dependence on radius would be expected from momentum flux conservation if mass entrainment is occurring in regions where the radius increases (e.g. \citetalias{croston_14a}). We assumed $\delta_{\rm inj} = 2.1$, consistent with the work of \citet{laing_02a} and our observed spectra in the inner parts of the source (Appendix~\ref{flux}).

Our modelling approach then involves considering a range of input
values for $\beta$, $\varv_{0}$ and $B_{0}$, and applying the advection model
as described in Section~\ref{sec:advection_model} below to determine an
output magnetic field and velocity profiles $B(z)$ and $\varv(z)$ (where
$z$ is distance along the tail) that minimize $\chi^{2}$. However,
there is an intrinsic degeneracy between the velocity and magnetic
field strength profiles along the tail if we wish to avoid the
traditional minimum energy assumption in which the CRE and magnetic
field energy densities are approximately equal with no relativistic protons. Hence
there are multiple combinations of parameters that can achieve a good
fit to the radio data. 

Our initial approach to tackling the degeneracy between velocity and
magnetic field strength is to compare the best-fitting
magnetic field profiles for a range of input parameters
($\beta$,~$\varv_{0}$,~$B_{0}$) with the expected magnetic field strength
profile of models that use further observational constraints -- specifically, assuming pressure equilibrium with the surrounding X-ray emitting ICM (e.g.\ \citetalias{croston_14a}) -- in order to assess their physical plausibility. In Section~\ref{sec:results}, we also discuss alternative scenarios.

\subsection{Advection model}
\label{sec:advection_model}

Our quasi-1D cosmic-ray transport model (based on \citetalias{heesen_16a}) is
able to predict non-thermal (synchrotron) radio continuum intensity
profiles at multiple frequencies, assuming advection is the dominant
transport process. For no in-situ cosmic-ray acceleration in the tails, the stationary
(no explicit time dependence) advection model prescribes the CRE flux
(units of $\rm s^{-1}\,GeV^{-1}$) $N(E,z)$ as:
\begin{equation}
  \frac{\partial N(E,z)} {\partial z} = \frac{1}{\varv}\left [
    \frac{\partial}{\partial E}\left ( b(E)
    N(E,z)\right )\right ]\qquad (z>0),
\label{eq:adv}
\end{equation}
where $z$ is distance along the tail, $E$ is the electron energy,
$\varv$ is the advection speed (which may depend on position) and $b(E)$ include the losses of the CREs via IC and synchrotron radiation as well as adiabatic losses:
\begin{eqnarray}
b(E) & = & -\left (\frac{{\rm d}E}{{\rm d}t}\right )= -\left (\frac{{\rm d}E}{{\rm
      d}t}\right )_{\rm IC} -\left (\frac{{\rm d}E}{{\rm d}t}\right )_{\rm syn}
-\left (\frac{{\rm d}E}{{\rm d}t}\right )_{\rm ad} \\
& = & \frac{4}{3} \sigma_{\rm T} c \left (\frac{E}{m_{\rm  e}c^2} \right )^2
U_{\rm rad} + \frac{4}{3} \sigma_{\rm T} c \left (\frac{E}{m_{\rm  e}c^2}
\right )^2 U_{\rm B} + \frac{E}{t_{\rm ad}},\nonumber
\end{eqnarray}
where $U_{\rm rad}=4.2\times 10^{-13}~\rm erg\, cm^{-3}$ is the radiation
energy density (here the only source is the CMB), $U_{\rm B}=B^2/(8\upi)$ is the magnetic
energy density, $\sigma_{\rm T}=6.65\times 10^{-25}~\rm cm^2$ is the
Thomson cross section and $m_{\rm e}=511~\rm keV\,c^{-2}$ is the electron
rest mass. Adiabatic losses are caused by the longitudinal dilution of the CRE
density due to an accelerating flow or due to lateral dilution by an expanding
flow. The corresponding losses of the CREs is ${\rm d}E/{\rm d}t= -1/3(\nabla
\varv)E$ \citep{longair_11a}. This means we can calculate the adiabatic loss time-scale as (Appendix~\ref{adiabatic}):
\begin {equation}
  t_{\rm ad} = \frac{3}{2}\left ( \frac{\varv}{r} \cdot \frac{\partial r}{\partial
      z} \right ) + 3\frac{\partial \varv}{\partial z},
\end{equation}
where $r$ is the tail radius. 
In \citetalias{heesen_16a}, $N(E,z)$ describes the CRE number density, which is identical to the CRE flux when the advection velocity and cross-sectional areas are constant. In our case, however, we need to convert the CRE flux into the CRE number density (units of $\rm cm^{-3}\, GeV^{-1}$):
\begin{equation}
  n = \frac{N}{A\cdot \varv}.
\end{equation}
Then, the \emph{model} emissivity can be calculated with the following expression using $A=\upi r^2$: 
\begin{equation}
   \epsilon_{\nu}  = {\rm const.} \int_0^\infty j(\nu) \frac{N(E,z)}{r^2 \varv} {\rm d}E,
\label{eq:emissivity}
\end{equation}
where $j(\nu)={\rm (const.)}B(z) F(\nu)$ is the synchrotron emissivity of a single
ultra-relativistic CRE \citep{rohlfs_04a}. Here, we have assumed a uniform distribution of the CRE number density and that the tails are cylindrical over small ranges of distances from the nucleus $\Delta z$. Now, we determine the \emph{observed} emissivity profiles. The emissivity $\kappa_{\nu}$ of a small volume in the tails can be calculated in a straightforward way as:
\begin{equation}
   \kappa_{\nu} = \frac{S_{\nu}}{\upi r^2\Delta z},
\end{equation}
where $\upi r^2\Delta z$ is the volume from which the emission stems, noting that $z$ is the physical distance from the nucleus. In order to convert the observed intensities into emissivities, we first calculate the flux density of the emitting volume:
\begin{equation}
S_{\nu} = \langle I_{\nu} \rangle \cdot \frac{2r\Delta z \cdot \sin(\vartheta)}{\Omega\cdot D^2},
\label{eq:flux_density}
\end{equation}
where $\langle I_{\nu} \rangle$ is the average intensity across the integration area, $\vartheta=52\degr$ is the inclination angle of the tail to the line of sight and $\Omega$ is the solid angle of the synthesized beam. We measure the radius $r$ of the tails from our 145-MHz low-resolution map and from our 360-MHz full-resolution map in the inner part. Then, the emissivity is:
\begin{equation}
\kappa_{\nu} = \langle I_{\nu} \rangle \cdot \frac{2\cdot \sin(\vartheta)}{\upi r \Omega\cdot D^2}.
\end{equation}
Combining all the constants and retaining only the variables, we find for the observed emissivity:
\begin{equation}
	\kappa_{\nu} = 1.16\times 10^{-3}\cdot \frac{\langle I_{\nu} \rangle}{r / {\rm kpc}}~{\rm Jy\, kpc^{-3}},
    \label{eq:observed_emissivity}
\end{equation}
The observed emissivity profiles are shown in Fig.~\ref{fig:alpha}.
\begin{figure*}
  \includegraphics[width=1.0\hsize]{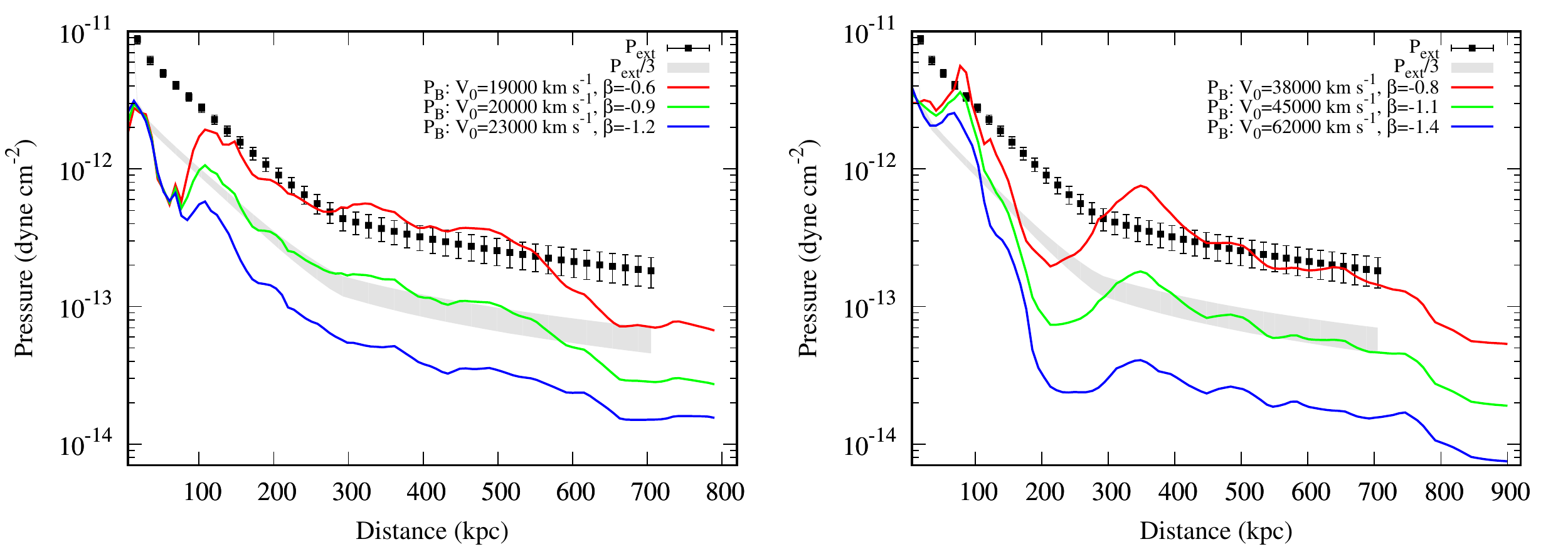}
  \caption{Comparison of the external pressure $P_{\rm ext}$ (data points)
    with the magnetic pressure $P_{\rm B}$ in the northern (\emph{left}) and southern tail (\emph{right}). The grey-shaded area indicates the range of pressures if the external
    pressure, extrapolated beyond the observed profile of \citetalias{croston_14a} extending to $\approx$$200$~kpc using their best-fitting model, is divided by a constant factor of $3$, which is the expected magnetic
    pressure for the entrainment model of \citetalias{croston_14a}  (see text for details). The various lines are the best-fitting advection models for Model~I ($B_0=15~\umu\rm G$) and for various values of $\beta$.}
\label{fig:pressure}
\end{figure*}
%\newpage
\begin{table}
\centering
\caption{Results of the cosmic-ray transport modelling.\label{tab:crtrans}}
\begin{tabular}{l cc}
\hline\hline
Parameter & Northern tail  & Southern tail\\\hline
Injection spectral index  ($\delta_{\rm inj}$) & \multicolumn{2}{c}{$2.1$}\\
Deceleration ($\beta$, $\varv\propto r^{\beta}$) & $-0.9$ & $-1.1$ \\
Degrees of freedom ($\rm dof$) & 83 & 67\\
\hline
\multicolumn{3}{c}{--- Model I: $B_0$ from CH14 entrainment model ---}\\
\hline
Injection magnetic field ($B_0$) & \multicolumn{2}{c}{$15~\umu \rm G$}\\
Injection velocity ($\varv_0$) & $20000~\rm km\, s^{-1}$ & $45000~\rm km\, s^{-1}$\\
Advection time ($\tau_{\rm adv}$)  & 190~Myr & 160~Myr\\
Reduced $\chi^2$ & $1.3$ & $1.6$\\
\hline
\multicolumn{3}{c}{--- Model II: $B_0$ assuming magnetically dominated
  tails ---}\\
\hline
Injection magnetic field ($B_0$) & \multicolumn{2}{c}{$26~\umu \rm G$}\\
Injection velocity ($\varv_0$) & $40000~\rm km\, s^{-1}$ & $80000~\rm km\, s^{-1}$\\
Advection time ($\tau_{\rm adv}$)  & 100~Myr & 90~Myr\\
Reduced $\chi^2$ & $1.3$ & $1.7$\\
\hline
\multicolumn{3}{c}{--- Model III: $B_0$ assuming equipartition and no
relativistic protons ---}\\
\hline
Injection magnetic field ($B_0$) & \multicolumn{2}{c}{$9~\umu \rm G$}\\
Injection velocity ($\varv_0$) & $13000~\rm km\, s^{-1}$ & $31000~\rm km\, s^{-1}$\\
Advection time ($\tau_{\rm adv}$)  & 300~Myr & 240~Myr\\
Reduced $\chi^2$ & $1.6$ & $1.4$\\
\hline
\end{tabular}
\end{table}

\citetalias{heesen_16a} developed the computer code {\small
  SPINNAKER} (SPectral INdex Numerical Analysis of K(c)osmic-ray
Electron Radio-emission), which implements the methods discussed above
to model the radio continuum emission.\footnote{{\scriptsize
    SPINNAKER} will be released to the public at a later
  date as free software.}  The right-hand side of equation~(\ref{eq:adv})
is solved numerically using the method of finite differences on a
two-dimensional grid, which has both a spatial and a frequency
(equivalent to the CRE energy) dimension. The left-hand side of
equation~(\ref{eq:adv}) can then be integrated from the inner boundary
using a Runge--Kutta scheme \citep[e.g.][]{press_92a}, which provides
us with a profile of $N(E,z)$. The initial parameters (see Section~\ref{sec:modelling_approach_and_assumptions}) are listed in Table~\ref{tab:crtrans}.

\subsection{Fitting procedure}

For each set of input parameters ($B_0$, $\varv_0$, $\beta$), we used
{\small SPINNAKER} to fit equation~(\ref{eq:emissivity}) to the
observed radio emissivity profiles obtained from equation~(\ref{eq:observed_emissivity}). We fit only to the 145-MHz data within distances $z<140~\rm kpc$ (in the northern tail) and $|z|<360$~kpc (in the southern tail) because the spectral index in this area becomes unreliable due to insufficient angular resolution (resulting in a superposition of different CRE populations), but include the other three radio frequencies beyond these distances. The reduced $\chi^2_{\rm red}$ is:
\begin{equation}
  \chi^2_{\rm red} = \frac{1}{\rm dof} \sum \left ( \frac{\kappa_{\nu,i} -
      \zeta \cdot {\epsilon_{\nu,i}}}{\sigma_i} \right )^2.
      \label{eq:reduced_chi_squared}
\end{equation}
Here, $\kappa_{\nu,i}$ is the $i$th observed emissivity and $\epsilon_{\nu,i}$ the
corresponding model emissivity, $\sigma_i$ the error of the observed value and $\rm dof$ is the degree of freedom (number of data points minus number of fitting parameters). The normalization parameter $\zeta$ is
determined by minimizing the reduced $\chi^2$. The magnetic field is
then varied with an iterative process where the 145-MHz observed emissivities are fitted, assuming that the model emissivities scale as $\epsilon_{\nu}\propto B^{1-\alpha}$. If the magnetic field has to be adjusted locally, the field has to change as $B\propto \epsilon_{\nu}^{1/(1-\alpha)}$ (which is accurate if $N(E,z)$ follows a power law in energy):
\begin{equation}
B_{j+1}(z)=B_{j}(z)\times \left ( \frac{\kappa_{145}(z)}{\xi
    \cdot [\epsilon_{145}(z)]_{j}} \right )^{1/(1-\alpha_{52-145})},
\label{eq:magnetic_field_fit}
\end{equation}
where $\alpha_{52-145}$ is the radio spectral index between 52 and
145~MHz. This is repeated several times until the fit converges and
the reduced $\chi^2$ does not decrease any further. Here, $j$ is an
integer variable, describing consecutive magnetic field models. While fitting equation~(\ref{eq:magnetic_field_fit}), we vary the normalization factor of the model emissivities $\xi$ so that $B_0$ is unchanged. In summary, the procedure is as follows:
\begin{enumerate}
\item Prescribe the magnetic field strength profile.
\item Fit the model, deriving the scaling $\zeta$ by minimize the reduced $\chi^2$ (equation~\ref{eq:reduced_chi_squared}).
\item Adjust the magnetic field profile using equation~(\ref{eq:magnetic_field_fit}), varying $\xi$ so that $B_0$ is fixed.
\item Iterate (ii) and (iii) until $\chi^2$ converges.
\end{enumerate}
With this procedure the 145-MHz emissivities are perfectly fitted since all the variations in the profile are absorbed by the magnetic field profile. The real test for our model is that the other frequencies are fitted as well, hence the spectral index profiles are most important.

\begin{figure*}
  \includegraphics[width=1.0\hsize]{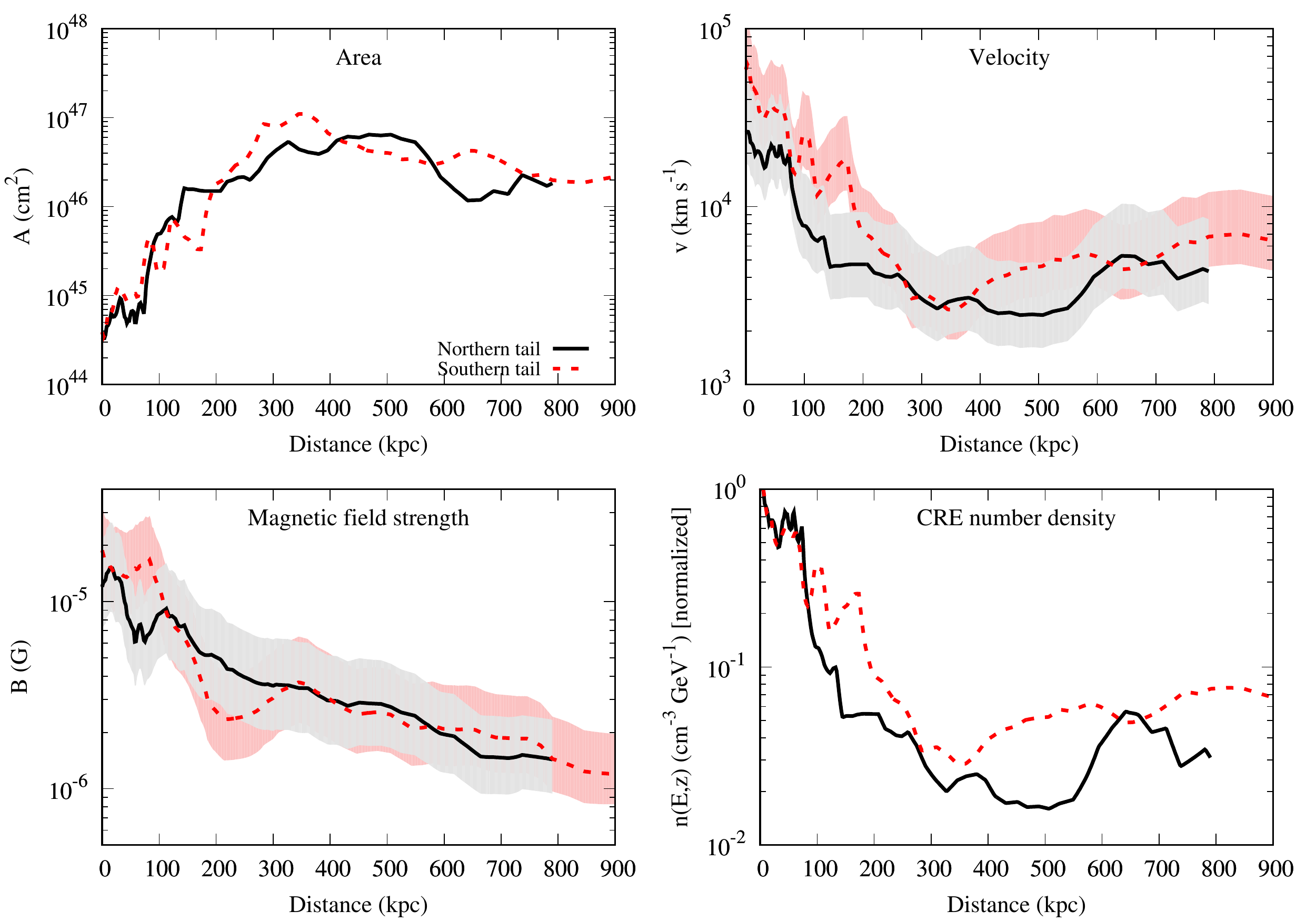}
  \caption{Cross-sectional area
    (\emph{top left}), velocity (\emph{top right}), magnetic field strength
    (\emph{bottom left}) and CRE number density corresponding to a critical
    frequency of 145~MHz (\emph{bottom right}) as function of distance to the
    nucleus. Profiles in the northern (southern) tail are shown as black solid
    (red dashed) lines, which are for Model~I. The shaded areas indicate the variation of velocity and magnetic field strength when using Models~II and III.}
\label{fig:val}
\end{figure*}
\subsection{Results}
\label{sec:results}
Table~\ref{tab:crtrans} lists the model fitting results for three
scenarios. In each case, we fixed $B_{0}$ and $\beta$, as explained
below, and then determined the best-fitting magnetic field profile and
$\varv_0$.

For our primary model (Model I in Table~\ref{tab:crtrans}), we require the tail to be in rough pressure balance with the external medium, as measured by \citetalias{croston_14a}. In their favoured model of dominant thermal pressure from entrained material, it was assumed that the magnetic field energy density is in rough equipartition with the total particle energy density. This model predicts that the magnetic field
pressure contribution, $P_{\rm B}$, relates to the external pressure, $P_{\rm ext}$, as  $P_{\rm B}=P_{\rm ext}/3$. In Fig.~\ref{fig:pressure}, we show the external pressure
profile inferred from the measured profile of \citetalias{croston_14a}, which extends to $\approx$$200$~kpc, extrapolating their best-fitting double-beta model to 700 kpc. The grey-shaded area indicates the predicted magnetic field pressure evolution in this model. 

The magnetic field profile is almost solely determined by the profile of the velocity along the radio tails, due the strong effect of adiabatic losses on the evolution of CRE number densities. In Fig.~\ref{fig:pressure} we show the effect of varying $\beta$ on the magnetic pressure profile: the best agreement with the extrapolated pressure profile of \citetalias{croston_14a} is for a value of $\beta \approx -1$. If we now vary the initial magnetic field strength $B_0$ but keep $\beta=-0.9$ (northern tail) and $\beta=-1.1$ (southern tail), the curves in Fig.~\ref{fig:pressure} will move up and down, and the corresponding best-fitting $\varv_0$ values will change, but the magnetic pressure profile shape does not change. Table~\ref{tab:crtrans} compares the results of three models: Model~I, using $B_{0}= 15~\umu$G, taken from \citetalias{croston_14a}, Model~II in which the tails are magnetically dominated and the magnetic pressure is roughly equal to the external pressure, which corresponds to the most extreme high-$B$ scenario, and Model III, in which we assume that $B_0$ is the equipartition magnetic field assuming no relativistic protons. We used the magnetic field
distribution indicated by the shaded area in Fig.~\ref{fig:pressure} (Model I) as an initial guess and
calculated the predicted radio intensities. For Models II and III, we rescaled this magnetic field profile to obtain the corresponding value of $B_0$. We can be confident that the true magnetic field pressure is somewhere in between these extremes. We have tested other starting conditions, e.g. uniform and exponential magnetic field profiles, and find that the final profiles of velocity and field
strength after iteration are reasonably independent of the initial
field strength profile for a given value of $\beta$. The best-fitting
profiles are shown in Fig.~\ref{fig:val}, with the range allowed for
the three models showed by the shaded areas. The corresponding
best-fitting parameters are shown in Table~\ref{tab:crtrans}.
\begin{figure*}
  \includegraphics[width=1.0\hsize]{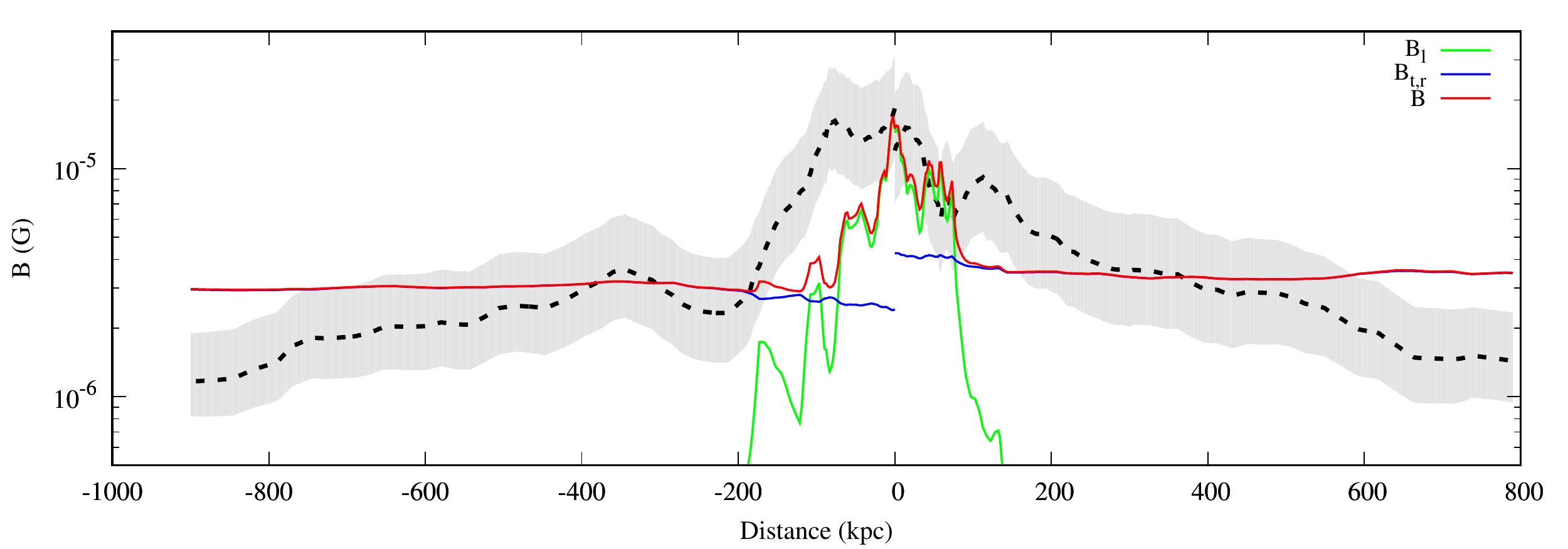}
  \caption{Adiabatic (flux conserving) magnetic field models. The black dashed line shows the best-fitting magnetic field strength from Model~I, where the shaded area indicates the variation when using Models II and III. The green and blue lines denote the best-fitting longitudinal and toroidal (radial) model components and the red line the total field strength. The northern tail is on the right-hand side ($z>0$~kpc) and the southern tail is on the left-hand side ($z<0$~kpc).}
\label{fig:bfield}
\end{figure*}

We find that our models describe the data reasonably well, with a
reduced $\chi^2\approx 1.5$. The spectral index profiles
(Fig.~\ref{fig:alpha}, bottom panel) show a continuous decrease
(steepening of the spectrum) with increasing slope with
distance. Notably, the shape of the spectral index profiles is
modelled accurately and has the typical shape of advective transport,
where the spectral index steepens in almost linear way with only a
slightly increasing curvature (cf.\ fig.~5 in
\citetalias{heesen_16a}). 

The resulting advection time-scale lies between 90 and 300~Myr. Our best-fitting velocity on scales exceeding $100$~kpc is broadly 
consistent with the estimated mean velocity of \citet{andernach_92a}, based
on data of considerably lower resolution and sensitivity. Our model assumes a steady-state flow, even in the outer parts of the tails, although in reality the tails are expected to be expanding in the longitudinal direction (the cylindrical shape in the outer few 100-kpc supports the assumption that the tails are not laterally expanding). We considered the effects of this growth on our assumptions, by comparing the time-scales for tail growth and advection of particles through the tails in the outer few 100-kpc. Since the cross-sectional area and the advection speed are both roughly constant in this region (Fig.~\ref{fig:val}), the fractional change in the length of a tail segment during the particle advection time-scale across that region is given by the ratio of the tail-tip advance speed and the advection speed. From ram-pressure balance (using the external pressure inferred from Fig.~\ref{fig:pressure}), which is an extrapolation from the inner model at the distance of interest, we estimate the tail-tip advance speed to be $\approx$$200~\rm km\,s^{-1}$, while the advection speed in the outer few 100-kpc is $\approx$$5000~\rm km\,s^{-1}$. Therefore the fractional increase in volume for a segment of jet over the particle advection time-scale is $\approx$4 per cent. As discussed in Section~\ref{sec:morphology_and_observed_spectrum}, it is likely that we are not detecting the full extent of the radio structure, and our external pressure estimate comes from an extrapolation rather than direct measurement at this distance. Consequently, the external pressure at the true end point of the source may be lower. However, the weak dependence of the advance speed on the external pressure means that this does not change the calculation very much for plausible environmental pressures. We conclude that the effect of tail growth on the particle energy evolution in the outer few 100-kpc should not significantly alter our results.

\section{Discussion}

\label{sec:discussion}

\subsection{Magnetic field structure}
\label{magnetic_field}
What regulates the magnetic field strength in the radio tails? One possibility
is simply that flux conservation is responsible. In cylindrical coordinates,
with the $z$-axis along the tail, the magnetic field has a longitudinal
component $B_{\rm l}$ and two transverse components, a radial $B_{\rm r}$ and a toroidal
$B_{\rm t}$ one. As \citet{laing_14a} have
shown, the magnetic field in FR~I jets to first order makes a transition from longitudinal
(first few kpc, where the jet is relativistic) to toroidal, occasionally with
a significant radial component. From flux conservation it is expected that the
longitudinal component falls much more rapidly than the two transverse
components because for a 1D non-relativistic flow \citep{baum_97a}:
\begin{equation}
\begin{array}{ll}
  B_{\rm l} \propto r^{-2}\quad {\rm (longitudinal)}\\[5pt]
  B_{\rm t},~B_{\rm r} \propto (r\varv)^{-1} = r^{-1-\beta} \quad {\rm (toroidal, radial)},
\end{array}
\label{eq:flux_conservation}
\end{equation}
where $r$ is the tail radius and $\varv$ the advection velocity. While we expect the radial component to fall off rapidly with radius
with $B_l\propto r^{-2}$, the toroidal and radial components
should be almost constant if
$\beta\approx -1$. We have fitted these models to our recovered magnetic
field distribution (Fig.~\ref{fig:val}), using the following normalization:
\begin{equation}
	B = B_{0} \cdot \left [ f^2 \cdot \left ( \frac{A}{A_0} \right )^{-2} + (1 - f^2) \cdot \left ( \frac{A}{A_0} \right )^{-1-\beta} \right ] ^{0.5},
\end{equation}
where $A_0$ is the cross-sectional area of the tail and $f$ is the ratio of  longitudinal to total magnetic field strength at $|z|=15$~kpc. The resulting flux-freezing magnetic field models are shown in
Fig.~\ref{fig:bfield}. At injection, the best-fitting flux conservation scenario requires that the longitudinal magnetic field component totally dominates and the radial and toroidal field strengths are very small. At distances $>$100~kpc from the nucleus, the longitudinal component becomes rapidly negligible compared with the toroidal and radial magnetic field components. However, the flux-freezing models do not fit the observed magnetic field profile very well. At best, our advection model results are consistent with such a flux conservation model out to a distance of $\approx$500~kpc. At larger distances, magnetic flux cannot be conserved as the resulting magnetic field pressure becomes too high. It is clear from the relatively flat profiles of cross-sectional
area and velocity shown in Fig.~\ref{fig:val} that the magnetic field evolution should be roughly constant if flux is conserved, whereas our model, strongly constrained by the observations, implies a decreasing magnetic field strength with distance. We note that similar geometrical arguments apply to the model of \citetalias{croston_14a} so that flux freezing is not consistent with the evolution of the inner plumes in their model. An alternative scenario is that the magnetic flux evolution is affected by entrainment, either continuously along the tails or at particular locations where some level of disruption occurs, with energy transferred between the magnetic field and evolving particle population. 

\subsection{Source age and energetics}
\label{source_age}
An important characteristic of radio galaxies is the source age, which
can be measured either as a dynamical time-scale if the source
expansion speed is known, or as a spectral time-scale derived from
the spectral ageing of the CREs \citep[e.g.][]{heesen_14a}. We can use
the sound crossing time as a first estimate for the source age
\citep{wykes_13a}. The ICM temperature is approximately $1.5$~keV
(\citetalias{croston_14a}, their table~1), resulting in a sound speed
of $600~\rm km\,s^{-1}$, implying a sound crossing time of
$\approx$$1000$~Myr. This would provide an upper limit to the source age were we confident about the inclination of the outer tails, but if the outer parts are more projected than we assume based on the inner jet, the true sound-crossing time could be higher. If, instead, we assume that we are observed the oldest plasma in the outer parts of the tails, then we obtain a source age of $\approx$$200$~Myr. If our advection time-scale gives a realistic estimate of the source age, then the average source expansion speed is
$\approx$$5$ times the speed of sound. As discussed in Section~\ref{sec:results}, ram-pressure balance arguments suggest that the current expansion speed is considerably lower than this, so that the overall source expansion at the present time is {\it not} supersonic. This apparent inconsistency could be explained if there is a weak internal shock somewhere beyond the detected emission, with back-flowing material forming an undetected cocoon of older plasma. If no such cocoon of material exists, then the advection speeds found by our model must be too high, presumably because of the effects of in-situ particle acceleration on the radio spectrum, which cannot be accounted for in our model. 

We briefly consider further the detectability of a large-scale diffuse cocoon in our observations. In Section~\ref{morphology}, we argued that our observations are not consistent with the claimed detection of a halo surrounding 3C~31 by \citet{wezgoviec_16a}. However, there still may be low-energy electrons which are not detected in the current observations. Assuming our 145-MHz $3\sigma$ threshold of $1.3~\umu\rm Jy\,arcsec^{-2}$, it would be possible for a diffuse halo of up to 15--20~Jy brightness to evade detection in our observations, depending on the assumed geometry. Based on a rough minimum energy calculation (assuming the total energy required to power the halo is $\approx$$4/3$ the internal energy), we estimate that it is energetically feasible for a jet of 3C~31's power to generate such an ellipsoidal halo surrounding the visible low-frequency structure, on a timescale of $\approx$1000~Myr. Hence, our observations cannot distinguish between the scenario where our modelled advection speeds are correct, and a large-scale cocoon of low-energy electrons exists, and the scenario in which the spectral age estimates are incorrect.

We conclude that our model can self-consistently describe the dynamics and energetics of 3C~31; however, dynamical considerations suggest that the advection time of the oldest visible material is likely to underestimate the true source age. This could be explained by the existence of a large repository of older, undetectable plasma, as discussed above, or by the effects of in-situ particle acceleration on the radio spectrum. An improved model would need to take into account the variation of the magnetic field across the width of the tails, small-scale variations due to turbulence, and, most challengingly, any in-situ particle acceleration on large scales. It has long been reported that the dynamical ages of FR~I sources appear significantly larger than their spectral ages \citep[e.g.][and references therein]{eilek_97a} -- our deep, low-frequency observations and detailed modelling do not resolve this discrepancy for 3C~31.

\subsection{Comparison with earlier work and uncertainties}
Our injection advection velocity for our favoured Model~I in the northern tail is $0.06$c; this is lower than the $0.14$c that \citet{laing_02a} found at 15~kpc from the nucleus. In the southern tail, our advection speed of $0.15$c is, however, in good agreement with their measurements. In the northern tail, only for the magnetically dominated Model~II we recover advection speeds with the correct value ($0.13$c). Although we note that in the model of \citet{laing_02a} the jet velocity changes as function of radius, e.g. decreasing from $0.16$c (on-axis) to $0.11$c (edge) and since our advection speeds are averages across the width of the tail we can not expect a perfect agreement. In comparison to \citetalias{croston_14a}, who use the injection velocities of \citet{laing_02a} as input parameter for the inner boundary condition, our velocities in the northern tail are lower by factors of $\approx$$2.5$ and $\approx$$1.5$ in the northern and southern tails, respectively. Scaling our velocities with these constant factors, we find agreement mostly within $0.2$~dex out to a distance of 100~kpc.

Our measurements hence agree roughly with previous observations, but we will discuss here some of the uncertainties that will affect our results. First, we note that we assume a constant inclination angle of the radio tails which may not hold in the outskirts of the tails. Because we are measuring only projected
distances, we potentially under- or overestimate the advection speeds. However,
the advection times are unaffected by this, since the CRE travelling distances
differ by the same factor. Another uncertainty is the inhomogeneity of the ICM, which maybe the cause of the many direction changes of the radio tails (\citetalias{croston_14a}). It is hence expected that the velocity changes in a way different from the simple parametrisation with radius that we assume. This may also account for some of the
variation of the magnetic pressure in comparison to the external pressure
(Section~\ref{magnetic_field}). We have already earlier mentioned the effects of particle acceleration, which could result in the spectral ages underestimating the true source age .

Our models also rely on the assumption of a power-law injection of the CREs. With our 
measurement of the the power-law radio continuum spectrum down to $30$~MHz we can measure the CRE spectrum to energies of $0.25$~GeV, corresponding to $\gamma \approx 500$. However, lack of spatial resolution and calibration uncertainties prevent us from measuring spatially resolved spectra at the lowest frequencies and so the assumption of a power law with the injection spectral index of $\delta_{\rm inj}=2.1$ can not be tested. Another strong assumption is that of a homogeneous magnetic field, where we do not take into account the radial dependence of the magnetic field strength or the turbulent magnetic field structure. \citet{hardcastle_13a} showed that in such a case we might underestimate the spectral ages and hence overestimate the advection velocities.

\section{Conclusions}

\label{sec:conclusions}

We have conducted LOFAR low-frequency radio continuum observations of the nearby FR~I radio galaxy
3C~31 between 30 and 178~MHz. These data were combined with VLA observations
between 290 and 420~MHz, GMRT observations at 615~MHz and archive WSRT data at
609~MHz. We have modelled these data with a quasi-1D cosmic-ray transport model for
pure advection using {\small SPINNAKER}, taking into account synchrotron
losses in the magnetic field and IC losses in the CMB. We included adiabatic
losses due to the expansion of the tails, while assuming a decelerating flow that
partially compensates for this. These are our main results:
\begin{enumerate}
\item We have shown that 3C 31 is significantly larger than was
  previously known; it now appears to extend at least 51~arcmin in the
  north-south direction, equivalent to $1.1$~Mpc at the distance of
  3C~31, which enables it to be classed as a GRG. Taking into account the
  bends in the radio tails, we can now trace the plasma flow for
  distances of 800--900~kpc in each direction.
\item The radio spectral index steepens significantly in the radio tails. The
  spatially resolved spectrum in the radio tails displays significant
  steepening and curvature, indicative of spectral ageing. The spectral index
  steepening can be well described by an advective cosmic-ray transport model.
\item We found that a decelerating flow in which
  $\varv\propto r^\beta$, with $\beta=-1.1\ldots -0.9$, is required to
  compensate partially for the strong adiabatic losses in the tails
  due to lateral expansion; however, the resulting velocity profile is
  relatively flat at distances $>$$200$ kpc.
\item We have constructed a self-consistent advection model in which
  the magnetic field strengths are in approximate agreement with
  pressure balance with the surrounding ICM, where the pressure
  supplied by the magnetic field is $1/3$ of the external pressure and
  the remainder is provided by a hot but sub-relativistic thermal gas
  (\citetalias{croston_14a}).
\item The derived source age of $\approx$200~Myr is smaller than the sound crossing
time of 1000~Myr, which is expected, and the average Mach number for the
expansion of the tails into the ICM is $\approx$5. If our model is correct, then we would predict the presence of a large-scale low surface brightness cocoon surrounding the observed tails, which could contain significant flux below our surface brightness sensitivity limit.
\item In the absence of in-situ particle acceleration, the derived spectral ages are firm \emph{upper} limits on the cosmic-ray lifetimes, even taking
into account uncertainties in the magnetic field strength. The magnetic field strength in the tails is mostly $<$$3~\umu\rm G$, so that IC losses are comparable to synchrotron losses of the CREs. Lowering the magnetic field strength further does not lead to a significant increase of the spectral age because IC radiation takes over as the dominant loss mechanism.
\end{enumerate}

3C~31 represents the first object where it has been possible to carry out such a detailed study using
low-frequency images at high spatial resolution, taking
advantage of the high quality of the LOFAR data. In future, the
goal is to expand this to other famous FR~I radio galaxies and to
samples extracted from the LOFAR surveys \citep{shimwell_17a}. The
first part is already in progress, with the study of objects such as
3C~449, NGC~315 and NGC~6251 under way.

\section*{Acknowledgements}
{\small We would like to thank the anonymous referee for their detailed, helpful comments that have greatly improved the paper. VH and JC acknowledge support from the Science and Technology Facilities
  Council (STFC) under grants ST/M001326/1 and ST/R00109X/1. MJH is supported by STFC grant
  ST/M001008/1. GJW gratefully acknowledges the
  receipt of a Leverhulme Emeritus Fellowship. We thank Joe Lazio
for assistance with the imaging of the GMRT data and the staff of the GMRT that made these
observations possible. GMRT is run by the National Centre for Radio
Astrophysics of the Tata Institute of Fundamental Research. The research
leading to these results has received funding from the European Research
Council under the European Union's Seventh Framework Programme (FP/2007-2013)
/ ERC Advanced Grant RADIOLIFE-320745. LOFAR, the Low Frequency Array designed
and constructed by ASTRON, has facilities in several countries, that are owned
by various parties (each with their own funding sources), and that are
collectively operated by the International LOFAR Telescope (ILT) foundation
under a joint scientific policy. The National Radio Astronomy
  Observatory is a facility of the National Science Foundation operated under
  cooperative agreement by Associated Universities, Inc. This research has made use of the NASA/IPAC Extragalactic Database (NED) which is operated by the Jet Propulsion Laboratory, California Institute of Technology, under contract with the National Aeronautics and Space Administration.}

\bibliographystyle{mnras}

{\scriptsize  \bibliography{fr1}
}

{\small This paper has been typeset from a \TeX/\LaTeX file prepared by the author.}

\appendix

\section{Flux scale and uncertainty}
\label{flux}

\begin{table}
\centering
\caption{Integrated flux densities $S_{\nu}$.\label{tab:flux}}
\begin{tabular}{lccccc}
\hline\hline
$\nu$ (MHz) & 3C~34 (Jy) & 3C~31 (Jy) & 3C~31 (Jy)$^{\rm a}$ & Ref.\\
\hline
$33.9 $ & $74.3\pm 2.2$ & $68.5\pm 2.1$ & $47.9\pm 1.4$ & This work\\
$36.9 $ & $68.8\pm 2.1$ & $66.4\pm 2.0$ & $46.2\pm 1.4$ & This work\\
$38.0 $ & $74.3\pm 7.4$ & $62.7\pm 6.4$ & $-$ & 1\\
$39.9 $ & $63.3\pm 1.9$ & $62.2\pm 1.9$ & $42.4\pm 1.3$ & This work\\    
$42.3 $ & $61.1\pm 1.8$ & $57.1\pm 1.8$ & $38.6\pm 1.2$ & This work\\    
$46.0 $ & $55.8\pm 1.7$ & $54.0\pm 1.7$ & $37.7\pm 1.1$ & This work\\    
$48.3 $ & $53.5\pm 1.6$ & $54.0\pm 1.6$ & $37.2\pm 1.1$ & This work\\    
$51.3 $ & $50.8\pm 1.5$ & $50.6\pm 1.6$ & $34.8\pm 1.0$ & This work\\    
$54.4 $ & $47.3\pm 1.4$ & $48.7\pm 1.5$ & $32.5\pm 1.0$ & This work\\    
$56.7 $ & $44.3\pm 1.3$ & $47.7\pm 1.5$ & $30.9\pm 0.9$ & This work\\    
$59.4 $ & $42.3\pm 1.3$ & $45.8\pm 1.4$ & $30.3\pm 0.9$ & This work\\    
$62.2 $ & $40.3\pm 1.2$ & $43.5\pm 1.3$ & $28.9\pm 0.9$ & This work\\    
$65.2 $ & $38.8\pm 1.2$ & $42.0\pm 0.9$ & $26.6\pm 0.8$ & This work\\
$74.0 $ & $34.9\pm 3.5$ & $27.8\pm 2.8$ & $-$ & 2\\
$116.7$ & $21.2\pm 0.6$ & $31.4\pm 0.9$ & $23.0\pm 0.7$ & This work\\    
$120.2$ & $20.4\pm 0.6$ & $30.0\pm 0.9$ & $22.2\pm 0.7$ & This work\\    
$123.7$ & $20.0\pm 0.6$ & $29.4\pm 0.9$ & $21.9\pm 0.7$ & This work\\    
$127.3$ & $19.4\pm 0.6$ & $28.7\pm 0.8$ & $21.4\pm 0.6$ & This work\\    
$130.8$ & $19.0\pm 0.6$ & $28.2\pm 0.8$ & $21.0\pm 0.6$ & This work\\    
$134.3$ & $18.6\pm 0.6$ & $27.9\pm 0.8$ & $20.9\pm 0.6$ & This work\\    
$137.8$ & $18.3\pm 0.6$ & $27.8\pm 0.8$ & $20.8\pm 0.6$ & This work\\    
$141.3$ & $17.8\pm 0.5$ & $27.0\pm 0.8$ & $20.3\pm 0.6$ & This work\\    
$144.8$ & $17.4\pm 0.5$ & $26.6\pm 0.8$ & $20.1\pm 0.6$ & This work\\	     
$148.4$ & $17.0\pm 0.5$ & $26.3\pm 0.8$ & $19.9\pm 0.6$ & This work\\	     
$151.9$ & $16.5\pm 0.5$ & $25.7\pm 0.8$ & $19.5\pm 0.6$ & This work\\	     
$155.4$ & $16.1\pm 0.5$ & $25.2\pm 0.8$ & $19.2\pm 0.6$ & This work\\	     
$158.9$ & $15.7\pm 0.5$ & $24.7\pm 0.7$ & $18.8\pm 0.6$ & This work\\            
$162.4$ & $15.3\pm 0.5$ & $24.3\pm 0.7$ & $18.6\pm 0.6$ & This work\\	     
$165.9$ & $14.9\pm 0.4$ & $24.0\pm 0.7$ & $18.3\pm 0.5$ & This work\\	     
$169.4$ & $14.7\pm 0.4$ & $24.0\pm 0.7$ & $18.2\pm 0.5$ & This work\\	     
$173.0$ & $14.3\pm 0.4$ & $23.5\pm 0.7$ & $17.9\pm 0.5$ & This work\\
$178.0$ & $13.0\pm 1.3$ & $18.3\pm 1.6$ & $-$ & 1\\     
$296.0$ & $9.6 \pm 0.3$ & $16.5\pm 0.5$ & $14.0\pm 0.4$ & This work\\            
$312.0$ & $9.0 \pm 0.3$ & $15.5\pm 0.5$ & $13.2\pm 0.4$ & This work\\
$325.0$ & $8.5 \pm 0.8$ & $13.5\pm 1.4$   & $-$ & 3\\
$328.0$ & $8.4 \pm 0.2$ & $15.0\pm 0.4$ & $12.6\pm 0.4$ & This work\\	     
$344.0$ & $8.2 \pm 0.2$ & $14.7\pm 0.4$ & $12.3\pm 0.4$ & This work\\	     
$392.0$ & $7.3 \pm 0.2$ & $13.1\pm 0.4$ & $11.2\pm 0.3$ & This work\\	     
$424.0$ & $6.9 \pm 0.2$ & $12.3 \pm 0.4$ & $10.6 \pm 0.3$ & This work\\            
$608.5$ & $-$ & $9.6 \pm 0.3$ & $8.6 \pm 0.3$ & This work\\
$750.0$ & $2.8\pm 0.3$ & $8.2\pm 0.2$ & $-$ & 1\\ 
$1400.0$ & $-$ & $5.4\pm 0.3$ & $-$ & 1\\ 
$2700.0$ & $-$ & $3.5\pm 0.4$ & $-$ & 4\\
$4850.0$ & $-$ & $2.1\pm 0.3$ & $-$ & 5\\
$10700.0$ & $-$ & $1.3\pm 0.1$ & $-$ & 4\\
\hline
\end{tabular}
\flushleft  {\bf Notes.} $^{\rm a}$Flux density of the bright inner tails (see text).\\
{\bf References.} 1: 3CRR sample by \citet{laing_80a}; 2: VLSS redux survey by
\citet{lane_12a}; 3: WENSS survey by \citet{rengelink_97a}; 4:
\citet{kuehr_81a}; 5: \citet{becker_91a}.
\end{table}
\begin{figure*}
  \includegraphics[width=0.8\hsize]{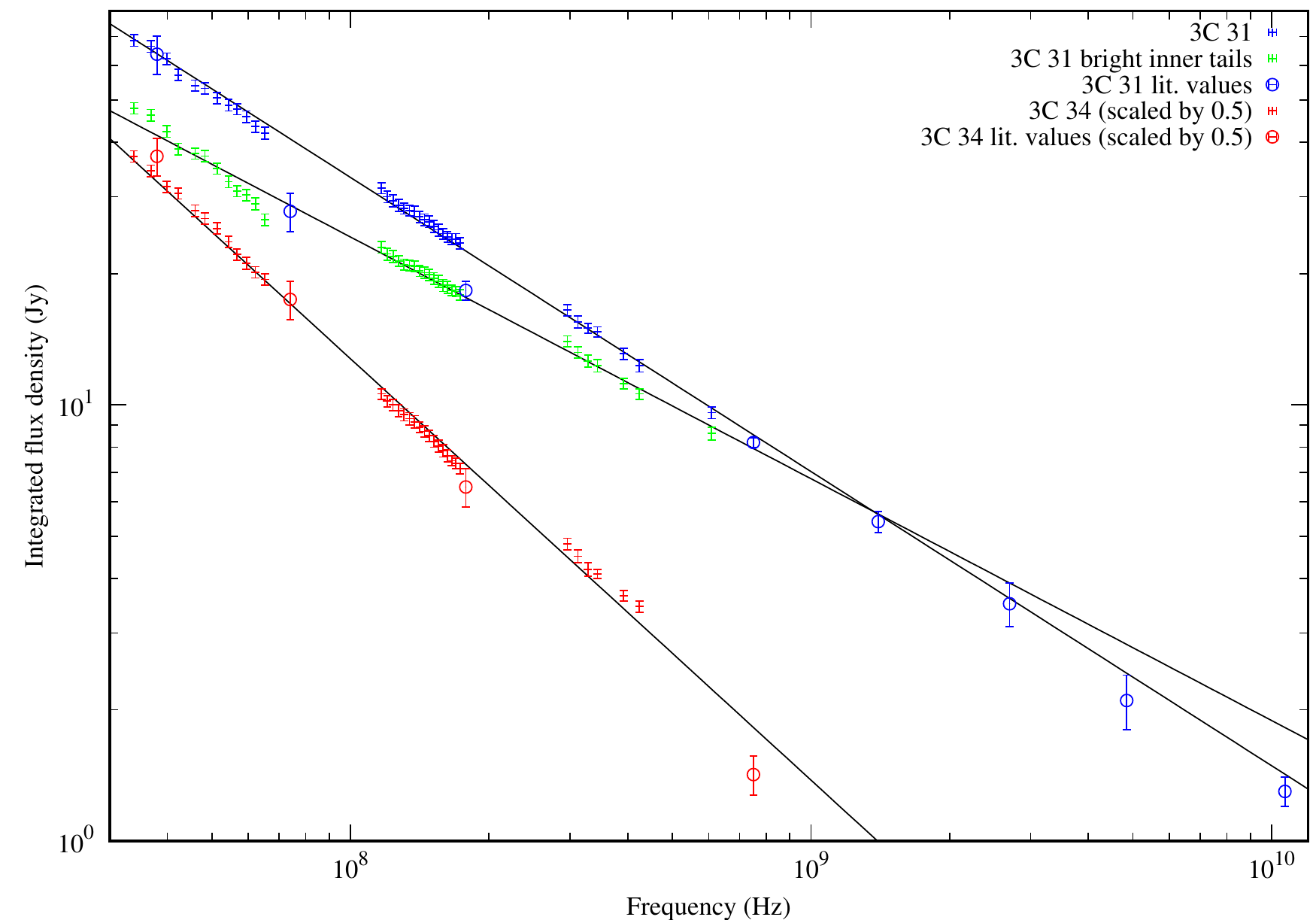}
  \caption{Spectrum of the integrated flux densities of 3C~31 and 34,
    where solid data points are our measurements and open data points are
    literature data. Lines show power-law fits to the data. The flux densities of 3C~34 were scaled by a factor of $0.5$ for clarity. The green data points show the flux densities of the
  bright inner tails of 3C~31 (see text for details).}
\label{fig:spectrum}
\end{figure*}

In this appendix, we present some more detail about the observations (see
Table~\ref{tab:observations} for a summary), in particular about the flux
densities, source spectra and uncertainties. Table~\ref{tab:flux} contains the integrated flux densities of 3C~31, our
science target, and of 3C~34, a compact, bright source located $0\fdg 9$
south-east of 3C~31. We used the {\small CASA}'s MS--MFS {\small CLEAN}ed LOFAR maps for 3C~31
and the {\small AWIMAGER} {\small CLEAN}ed LOFAR maps for 3C~34. {\small AWIMAGER} was
used for 3C~34, because it allows
for a correction for the LOFAR primary beam attenuation. The flux densities presented in this paper are scaled to
the flux scale of \citet*[][hereafter RCB]{roger_73a}. Our own observations
(LOFAR, VLA) were calibrated according to the calibrator models by
\citet{scaife_12a}, which utilize the RCB scale. The 178-MHz values presented by
\citet{laing_80a} were measured by \citet*[][hereafter
KPW]{kellerman_69a} and scaled with a factor of $1.09$ to the RCB scale. Similarly, the 38-MHz
value presented by \citet{laing_80a} was measured by KPW and scaled with a
factor of $1.10$ to the RCB scale. The 74-MHz values were obtained from the
VLSS redux survey, which are scaled to the calibrator models of
\citet{scaife_12a}. The 750-MHz values are from
\citet{pauliny-toth_66a} and were scaled by a factor of $1.046$ in
\citet{laing_80a}. The remaining values were scaled from the flux scale of
\citet*[hereafter B77] {baars_77a} to the RCB scale using the factors from
B77.\footnote{At frequencies $\nu > 325~\rm MHz$ the RCB scale
  agrees with the KPW scale \citep{scaife_12a}. So we used the conversion
  factors to scale from the B77 to the KPW flux scale.}

In order to check our flux scale, we investigated the spectrum of 3C~34 and
compared it with values from the literature. This source is particularly well suited to a
flux scale comparison
since it is unresolved in our observations at 1~arcmin resolution, and
similarly this is the case for the archive observations of the VLSS survey,
WENSS survey and other archive data, which all have resolutions around
1~arcmin.\footnote{Our VLA $P$-band observations of A- and B-array only
  (8~arcsec resolution) clearly resolve 3C~34 and
  show it to be a FR~II source with two hotspots separated by 40~arcsec (${\rm PA}=85\degr$).} In
Fig.~\ref{fig:spectrum}, we present our measured radio continuum fluxes of 3C~34
between 34 and 424~MHz obtained from our low-resolution maps. They can be well described by a power law with a radio
spectral index of $-0.97\pm 0.01$ and with a reduced $\chi^2$ of $2.1$. For the
error bars, we assumed a flux scale uncertainty of 3 per cent, which comes
solely from the uncertainty on the flux calibrator 3C~48
\citep{scaife_12a}. We find that the power-law fit to our data agrees
reasonably well with the literature values (shown as non-filled circles in Fig.~\ref{fig:spectrum}). 

For 3C~31, the
comparison with literature values is more complex: the reason is that the
source has faint and very extended radio tails, where the detection of emission is a function of
sensitivity. The spectrum of the integrated emission (inner part plus tails) in Fig.~\ref{fig:spectrum} can be described by a
power law, with a radio spectral index of $-0.67\pm 0.01$ between 34 and
$10700$~MHz. The literature data points at 74 and 178~MHz are too low, which deserves some
further investigation. The 178-MHz data point is based on the
KPW measurements using the Cambridge pencil-beam interferometer \citep[called
4CT pencil-beam by][]{laing_83a} which had a FWHM of 20~arcmin in north-south
orientation at the declination of 3C~31. As we shall see below, 3C~31 is essentially a source
of 51~arcmin extending in north-south direction at this frequency, so the KPW
178-MHz observations will certainly have resolved the source, but will have detected the bright tails as a point-like
source, picking up most of the flux. Similarly, the 74-MHz map of the VLSS
redux survey \citep{lane_12a} shows only the bright inner tails with little emission from the
tails. We can check the consistency of our data
points by simulating this effect, where we measured the flux density only in an area surrounding the bright inner tails (a rectangular box within R.A.\ $01^{\rm h} 07^{\rm m} 42^{\rm s}$--$01^{\rm h} 07^{\rm m}
  00^{\rm s}$, Dec.\ $+32\degr 18\arcmin$--$32\degr 33\arcmin$). These data are plotted in Fig.~\ref{fig:spectrum} as
  triangles; they can be fitted by a power law with a radio spectral index
of $-0.55\pm 0.01$ with reduced $\chi^2=2.2$ (the relatively high $\chi^2$
is due to the fact that the LBA in-band spectral index is too steep). Now the
best-fitting power law is in very good agreement with the KPW data
point. In the above measurements, we have assumed that the absolute calibration uncertainty is 3~per
cent, which is the uncertainty of the calibrator model of
\citet{scaife_12a}. This uncertainty is small, but is corroborated by our
finding of a reduced $\chi^2=1.0$ when fitting the integrated 3C~31 flux
densities (excluding the 74 and 178-MHz values) with a
power law.

Finally, we discuss the uncertainties of our spatially resolved
flux density
measurements. In the following analysis, we use our low-resolution maps, where
we average in a certain area (typically 3--10 beam areas) in order to find the
averaged intensity. There are three sources of uncertainty which we take into
account. Firstly, the calibration uncertainty $\rm cal_e=0.05$, which we
assume to be 5~per cent. It is slightly larger than the uncertainty of the
calibrator model, owing to the fact that imaging of extended emission results
into an increased uncertainty due to the imperfect $(u,v)$-coverage. Secondly,
there is the thermal noise in the $(u,v)$ visibilities, which results in approximately Gaussian
fluctuations of the intensities in the map quantified by $\sigma_I$. Thirdly, we find that the average
in the map is not always zero as expected for 
an interferometer. In areas of $\approx$100 beam areas, the average intensity
should be within $0.1\sigma_{\rm rms}$ with zero. But in our maps this
is not the case, the average intensity $\sigma_{\rm zero}$ is typically
comparable to $|\sigma_{\rm rms}|$ (so it can be either positive or negative). We add the contributions in quadrature, so
that we obtain for the error of the averaged intensities:
\begin{equation}
\sigma = \sqrt{ ({\rm cal_e} \cdot I_\nu)^2 + (\sigma_I/\sqrt{N_{\rm beam}})^2 + (\sigma_{\rm zero})^2}, 
\end{equation}
where $N_{\rm beam}$ is the number of beam areas within the integration
area.\footnote{Because this is an error estimate for intensities, one can divide the thermal
fluctuations by $\sqrt{N_{\rm beam}}$, rather than multiplying by
$\sqrt{N_{\rm beam}}$ as required for integrated flux densities.} In
Table~\ref{tab:rms}, we present the noise properties of our maps.

\begin{table}
\centering
\caption{Noise properties of our low-resolution maps (in units of $\rm mJy\,beam^{-1}$).
\label{tab:rms}}
\begin{tabular}{lcccc}
\hline\hline
 Map (MHz)& LBA (52) & HBA (145) & VLA (360) & WSRT (609)\\
\hline
$\sigma_{I}$ & 10 & $1.5$ & $1.0$ & $0.8$\\
$\sigma_{\rm zero}$ & 3 & $0.8$ & $0.5$ & $0.4$\\
\hline
\end{tabular}
\end{table}

\section{Adiabatic losses}
\label{adiabatic}

The   adiabatic losses (gains) are calculated as:
\begin{equation}
-  \left ( \frac{{\rm d}E}{{\rm d}t} \right )_{\rm ad}= \frac{1}{3} (\nabla
\varv)\cdot E = \frac{E}{t_{\rm ad}}.
\end{equation}
In cylindrical coordinates:
\begin{equation}
  \nabla \varv = \frac{1}{\rho} \frac{\partial \varv_{\rho}}{\partial \rho} +
  \frac{\partial \varv_{z}}{\partial z}.
\end{equation}
In the following, we assume that the tail expands (or contract) in a homologous
way, so that:
\begin{equation}
  \varv_{\rho}=\varv_{\rho 0}\cdot \frac{\rho}{r} \quad {\rm with} \quad \varv_{\rho
    0}=\frac{\partial r}{\partial t}=\frac{\partial r}{\partial z}\cdot \varv_{z},
\end{equation}
where $r$ is the tail radius. Hence:
\begin{equation}
  \nabla \varv = 2 \frac{\partial r}{\partial z}\cdot \frac{\varv_{z}}{r} +
  \frac{\partial \varv_{z}}{\partial z}.
\end{equation}
The adiabatic loss (gain) time-scale is then:
\begin{equation}
  \frac{1}{t_{\rm ad}}=\frac{1}{t_{{\rm ad}, r}} + \frac{1}{t_{{\rm ad}, \varv}},
\end{equation}
with the adiabatic losses (gains) through the tail lateral expansion
(contraction):
\begin{equation}
  t_{\rm ad, r} = \frac{2}{3} \frac{\partial r}{\partial z}\cdot
  \frac{\varv_{z}}{\partial z},
\end{equation}
and the adiabatic losses (gains) though the tail's velocity acceleration
(deceleration):
\begin{equation}
  t_{{\rm ad}, \varv} = \frac{1}{3} \frac{\partial \varv_{z}}{\partial z}.
\end{equation}
\begin{figure}
  \includegraphics[width=1.0\hsize]{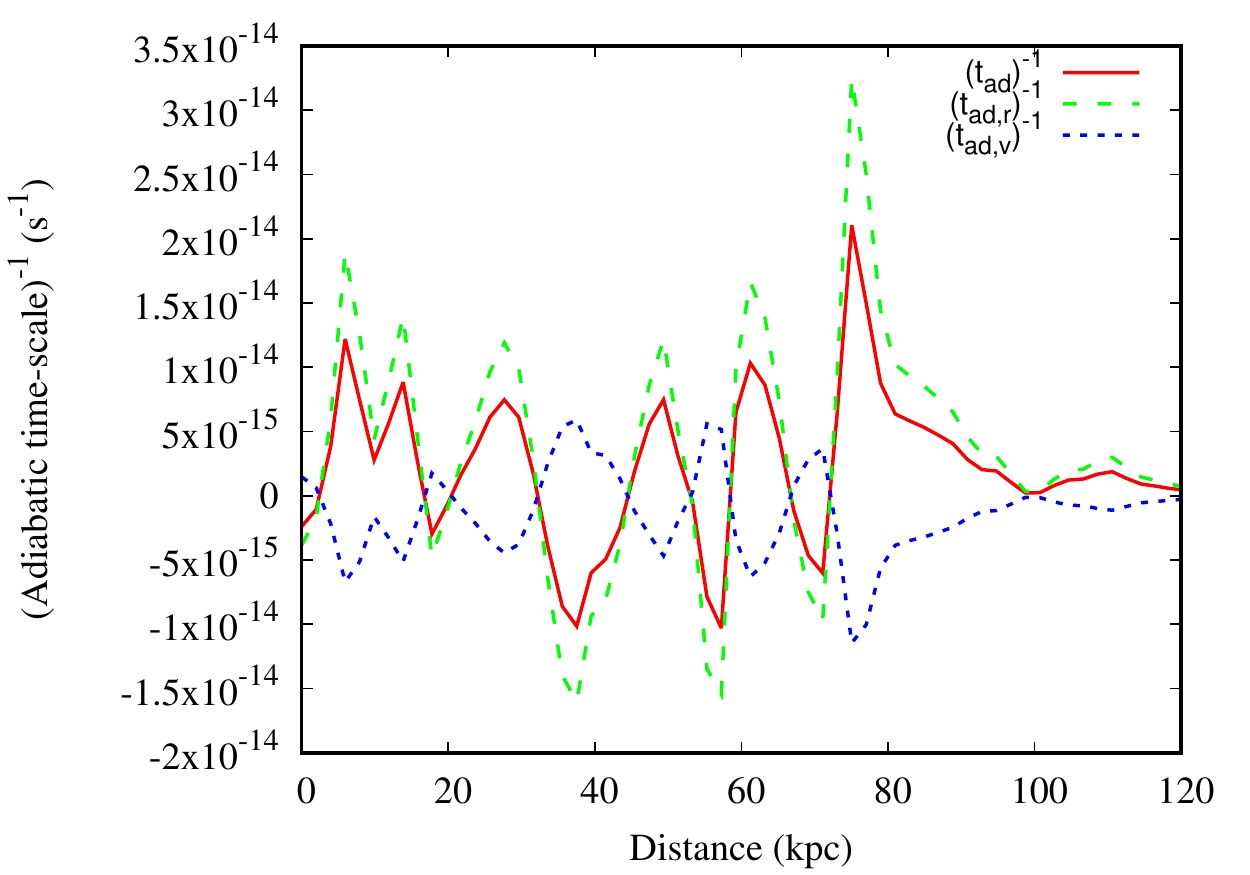}
  \caption{Inverse adiabatic time-scales in the northern tail with a
    decelerating flow ($\beta=-0.75$).}
\label{fig:tad}
\end{figure}
If the flow in the tail decelerates, the longitudinal compression 
cause adiabatic gains which offset partially the adiabatic losses due to the
lateral expansion. For a flow velocity $\varv\propto r^{\beta}$, with $\beta=-2$
the adiabatic losses vanish. For $\beta=-0.75$, the adiabatic losses are significantly
reduced as shown in Fig.~\ref{fig:tad}.
\begin{figure}
  \includegraphics[width=1.0\hsize]{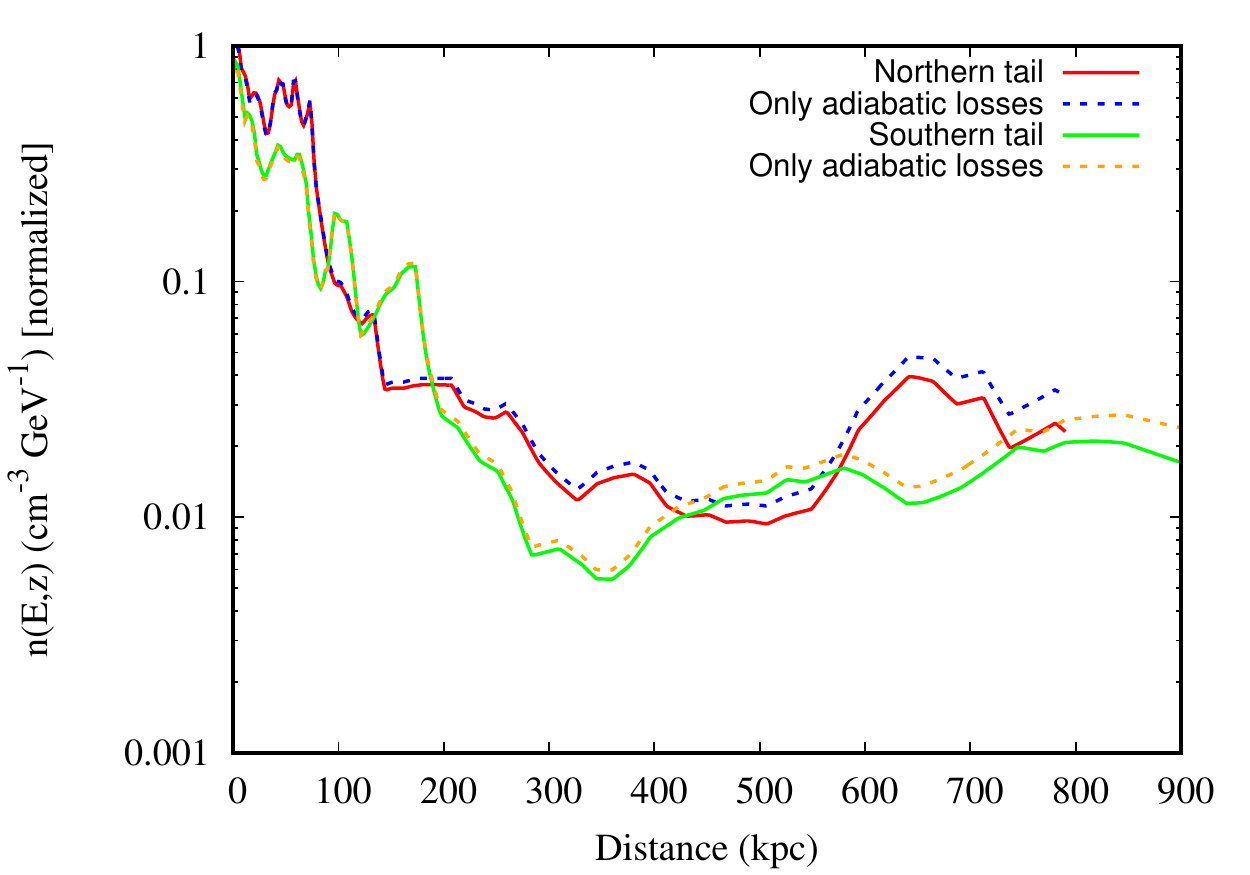}
  \caption{Comparison of the CRE (volume) number density at the critical frequency of 145~MHz with adiabatic models.}
\label{fig:nad}
\end{figure}
In order to test our numerical implementation, we compared our resulting CRE
number density with the analytical expression \citep[e.g.][]{baum_97a}:
\begin{equation}
  n \propto (\varv\cdot A)^{-(\delta_{\rm inj}+2)/3}.
\end{equation}
As can be seen in Fig.~\ref{fig:nad}, the analytical expression is in close
agreement with our numerical results. The small difference can be explained
that in our model the CREs have additional spectral losses via synchrotron and IC
radiation.

 \end{document}